\documentclass[12pt]{article}
		\title{Estimating the Sampling Distribution of Posterior Decision Summaries in Bayesian Clinical Trials}

\pdfoutput=1
\date{}
	\usepackage{amsmath,latexsym, graphics,graphicx,showexpl,amsthm,amsfonts,fullpage}
\usepackage[left=1in,top=1in,right=1in,nohead]{geometry}
\usepackage{graphicx}
\usepackage[margin=10pt,font={small},labelfont=bf]{caption}
\usepackage{subcaption}
\usepackage{bbm} 
\usepackage{setspace}
\usepackage{natbib}
\usepackage[titletoc]{appendix}
\usepackage{epstopdf}
\usepackage[pdftex,bookmarks,colorlinks=true,linkcolor=blue]{hyperref}
\usepackage{algpseudocode,algorithm}
\usepackage{authblk}
\usepackage{multicol}
\usepackage{hyperref}
\hypersetup{colorlinks,
	citecolor=blue
}
\usepackage{array}
\usepackage{etoolbox}
\usepackage{setspace}
\usepackage{rotating}

\newcolumntype{L}[1]{>{\raggedright\let\newline\\\arraybackslash\hspace{0pt}}m{#1}}
\newcolumntype{C}[1]{>{\centering\let\newline\\\arraybackslash\hspace{0pt}}m{#1}}
\newcolumntype{R}[1]{>{\raggedleft\let\newline\\\arraybackslash\hspace{0pt}}m{#1}}
\AtBeginEnvironment{thebibliography}{\linespread{1.5}\selectfont}

\makeatother
\author[1]{Shirin Golchi\thanks{shirin.golchi@mcgill.ca}}
\author[1]{James J. Willard}

\affil[1]{Biostatistics, McGill University, Montreal, QC H3A 1G1, Canada}
\begin{document}

	\maketitle

\begin{abstract}
Bayesian inference and the use of posterior or posterior predictive probabilities for decision making have become increasingly popular in clinical trials. The current practice in Bayesian clinical trials relies on a hybrid Bayesian-frequentist approach where the design and decision criteria are assessed with respect to frequentist operating characteristics such as power and type I error rate conditioning on a given set of parameters. These operating characteristics are commonly obtained via simulation studies. The utility of Bayesian measures, such as ``assurance", that incorporate uncertainty about model parameters in estimating the probabilities of various decisions in trials has been demonstrated recently. However, the computational burden remains an obstacle toward wider use of such criteria. In this article, we propose methodology which utilizes large sample theory of the posterior distribution to define parametric models for the sampling distribution of the posterior summaries used for decision making. The parameters of these models are estimated using a small number of simulation scenarios, thereby refining these models to capture the sampling distribution for small to moderate sample size. The proposed approach toward the assessment of conditional and marginal operating characteristics and sample size determination can be considered as simulation-assisted rather than simulation-based. It enables formal incorporation of uncertainty about the trial assumptions via a design prior and significantly reduces the computational burden for the design of Bayesian trials in general.
\end{abstract}

\noindent%

\section{Introduction}

Bayesian inference and decision making have become increasingly popular in the design and analysis of clinical trials, particularly in adaptive designs with stopping criteria, small early phase trials, or single arm trials where external information is utilized  \citep{berry_case_1993, spiegelhalter_bayesian_2003}. Posterior and posterior predictive probabilities are commonly used to make decisions and draw conclusions in Bayesian clinical trials \citep{berry_bayesian_nodate}. Despite the popularity of Bayesian methods in clinical trials, regulatory agencies require the designs and procedures to be assessed with respect to frequentist operating characteristics \citep{fdacderbent_adaptive_nodate}. The most common operating characteristics are classical power and type I error rate, as well as the probabilities of stopping early for efficacy or futility at interim analyses in adaptive designs. Evaluation of frequentist criteria requires understanding the sampling behaviour of the inference procedures. For Bayesian trials in particular, inference relies on the posterior distribution and the sampling distribution of posterior probabilities. Therefore, the current approach for using Bayesian methods in clinical trials is not ``fully Bayesian", but a combination of Bayesian and frequentist philosophies \citep{spiegelhalter_bayesian_2003, berry_bayesian_nodate}.

Sampling properties of posterior summaries used for decision making may be studied and evaluated based on the asymptotic properties of the posterior distribution itself. In fact, much of the Bayesian  sample size determination (SSD) literature takes advantage of these asymptotic properties to obtain approximate sample size or power estimates \citep{ohagan_bayesian_2001, kunzmann_review_2021, ohagan_bayesian_2001}. However, these approximations are poor for small sample sizes and improve at a slower rate for more complex and highly parametrized models. As a result, Monte Carlo simulations have been used to quantify frequentist operating characteristics of a proposed Bayesian design and analysis framework. \cite{gelfand_simulation-based_2002} formally introduced a simulation based approach toward Bayesian SSD. This approach involves iteratively simulating trials, with a particular design, analysis model and decision procedure, to estimate and control the probability of making erroneous decisions over the sampling distribution. 

Such simulation studies in modern trial design can involve a large number of scenarios, arising from various configurations of design parameters (e.g., sample size, number and spacing of interim analyses, randomization techniques), analysis model parameters (i.e., baseline, effect, and hyper/nuisance parameters) and decision parameters (i.e., efficacy and futility thresholds). The simulations are most often designed to mirror a frequentist power analysis. A simulation scenario is defined by assuming fixed values of the model parameters, with a focus on the effect parameter, and a given set of design and decision parameters. For each simulation scenario, data are generated from the model and the trial design is simulated for a large number of iterations (at least 10,000 iterations are recommended by the United States Food and Drug Administration --\cite{fdacderbent_adaptive_nodate}). If the posterior distribution arising from the specified statistical model is analytically intractable, each simulation iteration involves sampling or approximation techniques for estimating the posterior resulting in a significant computational burden. 

 It is argued that frequentist operating characteristics are inadequate for assessing the design of clinical trials, and more so for Bayesian designs, as they ignore the uncertainty at the design stage and condition on a fixed set of values of the parameters. Incorporating the uncertainty in the model parameters at the design stage has been extensively discussed in the Bayesian SSD literature. See \cite{kunzmann_review_2021} for a review. In particular, several authors have recommended the use of Bayesian assurance, an alternative measure to classical power, defined as the integrated or marginal probability of success over a ``design prior" on the parameter(s) of interest \citep{spiegelhalter_predictive_1986, ohagan_assurance_2005, chuang-stein_sample_2006, ren_assurance_2014}. More recently, the utility of assurance --also referred to as Bayesian predictive power or average power-- has been emphasized and demonstrated in practice within the context of trial design \citep{HamBorHol22, chapple_multi-armed_2023}. Obtaining marginal operating characteristics in Bayesian trials with non-conjugate, multi-parameter models is computationally intensive or even infeasible as it requires numerical integration of a function that can only be evaluated using time-consuming simulations.

As we explain in the following sections, the goal of this article is to propose methodology to efficiently execute many of the existing approaches in the Bayesian SSD literature for realistically complex models and designs. The proposed approach builds upon the theoretical and Monte Carlo-based approaches of Bayesian SSD in order to assess conditional and marginal operating characteristics of Bayesian analysis and decision procedures across the design/decision parameter space. This is achieved by taking advantage of the known asymptotic behaviour of the posterior distribution to define a relatively simple parametric model for the sampling distribution of posterior probabilities whose parameters are estimated from Monte Carlo simulations at a small number of selected scenarios. While we do not advocate for the use of conditional operating characteristics for the assessment of Bayesian designs, we are aware that these measures continue to be used in practice. Therefore, providing methodology that can assess these criteria allows for wider use of Bayesian designs by practitioners.

Previous work of a similar nature include \cite{muller_optimal_1995}  and \cite{muller_simulation_2005}, who propose a curve fitting approach to a target criterion, such as a utility or loss function, for Bayesian sample size determination; \cite{HAT_2024} consider the sensitivity of operating characteristics to simulation scenarios and propose methods to select the optimal set of scenarios to summarize the operating characteristics; \cite{golchi_estimating_2022} proposed a modelling approach for estimating the operating characteristics in Bayesian adaptive trials which was exclusively based on an initial set of simulations to estimate the sampling distribution. The present approach is different in that it combines asymptotic theory with simulations to learn the sampling distribution rather than target specific functionals of the distribution or rely solely on simulations. 

The methods proposed in this article can be viewed as a Bayesian power analysis approach that does not rely on large sample size assumptions or a large number of simulation scenarios. Such an approach is specifically beneficial for clinical trials where a complex analysis model is required and uncertainty about the model parameters needs to be incorporated into the power analysis. Considering the increasing complexity of research questions addressed in modern clinical trials, methodology that enables computationally efficient design of trials with multi-level models and high-dimensional parameter spaces is of increasing need. We anticipate that the present work will make a major contribution to the statistical design of modern clinical trials.

The remainder of this article is organized as follows. 
The problem setting and the proposed approach are presented in Section \ref{Sec:methods}. Section \ref{Sec:appl} provides a motivating example in the form of a hypothetical design exercise for an adaptive design with covariate adjustment where the performance of the approach and its applicability in estimating marginal operating characteristics is demonstrated. The article is concluded with a discussion in Section \ref{Sec:dis}.

\section{Methodology}
\label{Sec:methods}

Consider a clinical trial where the analysis model is defined using a set of parameters denoted by $\boldsymbol{\theta} \in \Theta$. Suppose that the research hypotheses are formulated as 
\begin{equation}
	H_0: \psi(\boldsymbol{\theta})\leq \psi_0 \hskip 10pt \text{vs} \hskip 10pt  H_A: \psi(\boldsymbol{\theta})>\psi_0
	\end{equation}
where $\psi(\boldsymbol{\theta})$ is the scalar target of inference referred to as the ``effect" parameter. 

Within a Bayesian analysis framework, a prior distribution is assigned to $\boldsymbol{\theta}$. In the context of clinical trials, analysis priors are often specified to be non-informative or weakly informative about the effect parameter. Informative priors may be used to reflect prior knowledge or external information about components of $\boldsymbol{\theta}$ that do not contribute to $\psi(\boldsymbol{\theta})$, referred to as the nuisance parameters. Inference is then carried out based on the posterior distribution 
$$\pi\left(\psi(\boldsymbol{\theta})\mid \mathbf{y}\right),$$ where $\mathbf{y}$ denotes the data.
For simplicity of presentation, we drop $\boldsymbol{\theta}$, which includes the nuisance parameters, from the notation in the remainder of the manuscript. 

Decisions and conclusions are then made based on the above posterior distribution using a decision summary, $\tau(\mathbf{y})$, which is commonly defined as the posterior probability that the alternative hypothesis is true. If this probability exceeds a predetermined threshold, $u$, the results are deemed conclusive:
\begin{equation}
	\label{eqn:pAlt}
	\tau(\mathbf{y}) = \pi\left(\psi>\psi_0\mid \mathbf{y}\right)>u.
	\end{equation}
Although the analysis may be performed within a Bayesian framework, the statistical design and decision procedures commonly follow the Neyman-Pearson hypothesis testing, where type I and II errors are controlled/optimized with respect to the sampling behaviour of the decision summary. In other words, while $\tau(\mathbf{y})$ is obtained by conditioning on the data, it is then treated as a function of the data, and its distribution over all possible trial data that could have arisen under the data model is of interest. 

Classical power is defined as the conditional probability of success given a hypothesized effect size, $\psi_d$,  under the alternative hypothesis, and fixed values of the nuisance parameters, although we omit the nuisance parameters from the notation for simplicity, $$
	P(\tau(\mathbf{y})>u\mid \psi_d). $$
Current Bayesian trial design most commonly relies on estimating this probability via Monte Carlo simulations, i.e., by repeatedly sampling data from the data model $\pi(\mathbf{y}\mid \psi=\psi_d)$, evaluating $\tau(\mathbf{y})$, and computing the sum $\sum_i \mathbbm{1}(\tau_i>u)$. 

The procedure described above is purely frequentist, in that it assumes a fixed true value for the parameters. However, the concept of power has been generalized within the Bayesian literature to incorporate uncertainty about $\psi_d$ in the power analysis. A Bayesian alternative measure to classical power is referred to as assurance \citep{ spiegelhalter_predictive_1986, ohagan_bayesian_2001, ohagan_assurance_2005, chuang-stein_sample_2006, ren_assurance_2014} which is the expected power with respect to a prior $\pi_d(\psi)$:
\begin{equation}
		\text{BA} = \text{E}_\psi \left[P\left(\tau>u\mid\psi\right)\right]=   \int P\left(\tau(\mathbf{y})>u\mid \psi\right)\pi_d(\psi)d\psi\label{BA}
	\end{equation}
Note that this prior distribution $\pi_d(\psi)$ is different from the prior distribution used in the analysis for a given trial. These two priors have been referred to as the \emph{design} versus \emph{analysis} priors, or \emph{sampling} versus \emph{fitting} priors, in the Bayesian SSD literature \citep{ohagan_bayesian_2001, gelfand_simulation-based_2002, pan_unifying_2021}. Classical power is a special case of assurance when the design prior is specified as a point mass. As discussed earlier, the computational cost of calculating marginal measures like assurance is prohibitively expensive, which is one reason why conditional operating characteristics are commonly used in practice.

To provide a solution, our goal is to obtain a parametric approximation of the sampling distribution of the posterior probability, $\tau$, whose parameters are estimated as functions of $\psi$ and the sample size, $n$, based on a number of simulated distributions at select values of $\psi$ and $n$. This parametric approximation can then be used to obtain Bayesian assurance by integrating over any design prior. 

Following \cite{golchi_estimating_2022}, we use the beta family where the parameters are in turn assigned probability distributions. As a result, the model for the sampling distribution is an infinite mixture of beta distributions weighted according to the distribution over the parameters. Instead of using Gaussian processes for the beta parameters as was proposed in \cite{golchi_estimating_2022}, we use parametric models that capture the known asymptotic behaviour of the posterior probability under the null and alternative hypotheses.

Consider $\psi = \psi_0$ as the point under $H_0$ where the type I error rate is maximized (i.e., the boundary of the null and alternative sets). Our interest is in estimating (controlling) the type I error rate at $\psi_0$. Suppose that a unique maximum likelihood estimator (MLE) exists for $\psi$, denoted by $\hat{\psi}_n$. Under weak regularity conditions (i.e., compact parameter space, continuity and identifiablity) the symmetry below follows from the Bernstein-von Mises theorem as well as asymptotic normality and consistency of the MLE \citep{vaart_asymptotic_1998},
\begin{equation}
	\label{eqn:von-Meses}
\psi\mid \hat{\psi}_n \overset{.}{\sim} \mathcal{N}(\hat{\psi}_n, \frac{1}{n}I^{-1}_{\hat{\psi}_n}), \end{equation}
\begin{equation}
	\label{eqn:CLT}
\hat{\psi}_n\mid \psi = \psi^* \overset{.}{\sim} \mathcal{N}( \psi^*, \frac{1}{n}I^{-1}_{\psi^*}).
\end{equation}
where $I^{-1}_{\hat{\psi}_n}$ and $I^{-1}_{\psi^*}$ denote the inverse Fisher information at $\hat{\psi}_n$ and $\psi^*$, respectively. Since the MLE is a function of the sufficient statistic for $\psi$, we can write the decision summary as 
\begin{align*}
	\tau &= P(\psi>\psi_0\mid \hat{\psi}_n)\\
	&=P(I^{1/2}_{\hat{\psi}_n}\sqrt{n}(\psi -  \hat{\psi}_n)>I^{1/2}_{\hat{\psi}_n}\sqrt{n}(\psi_0 -  \hat{\psi}_n)),
\end{align*}
which asymptotically yields the following from (\ref{eqn:von-Meses})
\begin{equation*}
	\tau \approx 1 - \Phi(I^{1/2}_{\hat{\psi}_n}\sqrt{n}(\psi_0 -  \hat{\psi}_n)) 
	= \Phi(I^{1/2}_{\hat{\psi}_n}\sqrt{n}( \hat{\psi}_n - \psi_0)),
\end{equation*}
where $\Phi$ is the standard normal cumulative distribution function (cdf). Then, by Slutsky's theorem (replacing $I^{1/2}_{\hat{\psi}_n}$ with $I^{1/2}_{\psi_0}$), we have
\begin{equation*}
	\tau\approx \Phi(I^{1/2}_{\psi_0}\sqrt{n}( \hat{\psi}_n - \psi_0)),
	\end{equation*}
which from (\ref{eqn:CLT}) is the cdf of a standard normal random variable and therefore follows a $U(0,1)$ distribution.

Similarly, given a point under the (separable) alternative, $\psi = \psi^*$, where there exists $\epsilon$, such that $||\psi^*-\psi_0||\geq\epsilon$, from (\ref{eqn:von-Meses}) we have
\begin{align*}
	\tau &\approx \Phi(I^{1/2}_{\hat{\psi}_n}\sqrt{n}( \hat{\psi}_n - \psi_0)) \\
	&= \Phi(I^{1/2}_{\hat{\psi}_n}\sqrt{n}( \hat{\psi}_n - \psi^* - (\psi_0 -\psi^*))),
	\end{align*}
resulting in the following by applying Slutsky's theorem (replacing $I^{1/2}_{\hat{\psi}_n}$ with $I^{1/2}_{\psi^*}$),
	\begin{align*}
		\tau&\approx \Phi(I^{1/2}_{\psi^*}\sqrt{n}( \hat{\psi}_n - \psi^*) - I^{1/2}_{\psi^*}\sqrt{n}( \psi_0- \psi^*)) \\
	&=\Phi( I^{1/2}_{\psi^*}\sqrt{n}( \hat{\psi}_n - \psi^*)+ I^{1/2}_{\psi^*}\sqrt{n}( \psi^*- \psi_0)),
\end{align*}
whose distribution converges to a point mass at 1 since the first term within the parentheses is asymptotically a standard normal random variable by (\ref{eqn:CLT}) and the second term grows with $\sqrt{n}$ since $( \psi^*- \psi_0)>0$.

Using the results above, we define parametric models for the sampling distribution of $\tau$ such that it is asymptotically a $U(0,1)$ given $\psi = \psi_0$ and a point mass at 1 given $\psi = \psi^*>\psi_0$. The parameters of these models will then be estimated from Monte Carlo simulated instances of the sampling distribution. 

Under the null hypothesis, it is often reasonable for the sampling distribution of $\tau$ to be approximately symmetric. The symmetry arises from the assumption that the posterior median is an unbiased estimator for $\psi$, whose true value under the null is assumed to be $\psi_0$ (See Appendix~\ref{sec:nullSymmetry}). Therefore, we specify a beta distribution with the same shape and scale parameters,
\begin{equation}
	\label{eq:betaH0}
	\tau\mid \psi = \psi_0 \sim \text{beta}\left(a_0(n), a_0(n)\right)
\end{equation}
where $n$ is the sample size and $a_0(n)>0$ is assigned a distribution indexed by $n$ such that as $n$ gets large the distribution above converges to a ${\rm U}(0,1)$. 
The following log-normal distribution on $a_0$ is defined to mimic the evolution of the sampling distribution as $n$ grows,
\begin{equation}
	\label{eq:aH0}
	\log\left(a_0(n)\right) \sim \mathcal{N}\left(\frac{\alpha_1}{n} + \frac{\alpha_2}{n^2}, \sigma^2_0\right)
\end{equation}
where $\alpha_1$ and $\alpha_2$ are estimated from the ``data", i.e., the simulated sampling distribution at select $n$. The role of $\alpha_1$ and $\alpha_2$ is to adjust the speed of convergence of the sampling distribution to a uniform distribution which depends on the complexity of the analysis model. In Appendix \ref{sec:asympmoments} we show that this specification asymptotically guarantees the first two moments of the uniform distribution. We note that having fixed the mean of the sampling distribution at 0.5, we expect the variance of the distribution to change with $n$. Therefore, it is reasonable to assume that the distribution converges to a uniform distribution in $1/n$ while the second term $\alpha_2/n^2$ can be considered a correction term enabling better fit. The variance parameter $\sigma^2_0$ accounts for the uncertainty in the beta parameters that are estimated from Monte Carlo samples of the distribution of $\tau$ as well as for the fact that a single beta distribution cannot capture the sampling distribution for finite $n$.

Given a point under the alternative hypothesis, $\psi=\psi^*$, and assuming $\psi^* - \psi_0>0$ without loss of generality, the sampling distribution of $\tau$ is modeled as follows,
\begin{equation}
	\label{eq:betaHA}
	\tau\mid \psi = \psi^* \sim \text{beta}\left(a_A(\sqrt{n}(\psi^* - \psi_0)), \frac{1}{a_A(\sqrt{n}(\psi^* - \psi_0))}\right)
\end{equation}
where the shape and inverse scale parameters depend on the distance from the null scaled by the square root of sample size. The choice of a $\text{beta}(a,b)$ with $b = 1/a$ is made for simplification and is not necessary. The scale parameter may be modelled with a different set of parameters. The only conditions that need to be met are that $a>b$ so that the distribution is left-skewed and that $a\rightarrow \infty$ and $b\rightarrow 0$ as $n$ gets large. As $\sqrt{n}(\psi^* - \psi_0)$ gets large, the sampling distribution is expected to become further skewed with a peak at 1, resembling a beta distribution with a large shape parameter and a small scale parameter converging to a point mass at 1 in the limit. The parameters are modeled such that the limiting distribution is guaranteed (Appendix \ref{sec:asympmoments}),
\begin{equation}
	\label{eq:aHA}
	\log\left(a_A\left(\sqrt{n}(\psi^* - \psi_0)\right)\right) \sim \mathcal{N}\left(\phi_1\sqrt{n}(\psi^* - \psi_0) + \phi_2 n(\psi^* - \psi_0)^2, \sigma^2_1\right)
\end{equation}
where $\phi_1$ and $ \phi_2$, similar to the $\alpha$'s above, are estimated from instances of the simulated distribution of $\tau$ at select $\psi^*$ and $n$, and adjust for varying rates of convergence of this distribution for different analysis models.

The sampling distribution is, therefore, approximated using an infinite mixture of beta distributions
\begin{equation}
	\label{eq:mix}
	f(\tau) \approx \int \omega(a)p(\tau;a)da
\end{equation}
 where $\omega(a)$ is given by the log-normal specifications and $p(\tau;a)$ by the beta distributions under either hypothesis above. Since the end goal is to estimate the operating characteristics as the tail probabilities of this distribution, this approximation only needs to capture the relevant portion of the sampling distribution. The relevant tail probabilities will depend on the range of the decision threshold $u$, which motivates the quantile matching approach that is described in the following \citep{sgouropoulos_matching_2015}. 
 
The beta mixture model described above is fit to simulated instances of the sampling distribution in two stages. For a given $n$ under the null hypothesis and for a pair of values $\psi^*$ and $n$ under the alternative hypothesis, let $q^e_1< \ldots < q^e_J$ be $J$ quantiles of the empirical distribution of $\tau$ generated via Monte Carlo simulations, and $q^{a}_1< \ldots < q^{a}_J$ be the corresponding theoretical quantiles of a beta distribution using either of the parametrizations presented above. In the first stage, an approximation for the sampling distribution is obtained as a member of the beta family whose upper quantiles best match the empirical upper quantiles. This is done via Bayesian least squares, i.e., by minimizing the sum of scaled squared loss through the general Bayesian framework of \cite{bissiri_general_2016}, which is equivalent to making inference via the following posterior distribution,

\begin{equation}
	\label{eqn:postab}
	\pi(a\mid \mathbf{q}^e) \propto \exp\left(-\frac{1}{\sigma_\epsilon}\sum_{j = 1}^J (q^e_j - q^{a}_j)^2\right) \pi(a),
	\end{equation}
where $\pi(a)$ is a weakly informative prior. 

Then samples drawn from the posterior distribution in (\ref{eqn:postab}) are used as ``data'' for the second stage to fit the models in (\ref{eq:betaH0})-(\ref{eq:aHA}). The second stage fit is done within the Bayesian framework by assigning weakly informative priors to the parameters $\alpha$'s, $\phi$'s and $\sigma$'s.  

To summarize, consider the sampling distribution under the alternative that is specified by the parameters $\boldsymbol{\phi} = (\phi_1, \phi_2)$ and $\sigma_1$. The posterior of interest is $\pi(\boldsymbol{\phi}, \sigma_1 \mid Q)$, where $Q_{D\times J}$ is the matrix of empirical quantiles obtained at $D$ simulation scenarios. To sample from 
\begin{equation}
\pi(\boldsymbol{\phi}, \sigma_1 \mid Q) = \int \pi(\boldsymbol{\phi}, \sigma_1 \mid a, Q)\pi(a\mid Q)da
\end{equation}
 we first sample from $\pi(a\mid \mathbf{q}^e_d)$ in (\ref{eqn:postab}) for $d = 1, \ldots, D$, and use these samples to estimate $ \pi(\boldsymbol{\phi}, \sigma_1 \mid a, Q)= \pi(\boldsymbol{\phi}, \sigma_1 \mid a)$ with the likelihood arising from (\ref{eq:aHA}).

\section{Application to trials with covariate adjustment}
\label{Sec:appl}

In this section, we illustrate the proposed methods within the context of covariate adjustment in clinical trials. While it is common practice to ignore prognostic information in the primary analysis of clinical trials data on the account of randomization having been performed, the benefits of covariate adjustment in clinical trials has been recently emphasized by many authors \citep{benkeser2021improving, van2022combining, lee2022benefits, willard_covariate_2022}. In particular, \cite{willard_covariate_2022} emphasize the advantages of adjusting for covariates in Bayesian adaptive clinical trials. Incorporating covariates in the primary analysis requires the simulations performed at the design stage to also include these covariates (with assumed distributions and effects).

As a simple example, consider a two-arm clinical trial design, where the vector of dichotomous primary outcomes is $\mathbf{Y} \sim {\rm Benoulli}(\mathbf{p})$ and where the alternative hypothesis specifies that the intervention reduces the probability of event. Let $\mathbf{X}$ be the design matrix which includes any prognostic covariates. A logistic regression model can then be used for the primary analysis,
\begin{equation}
	\label{eqn:adj}
	{\rm logit} (\mathbf{p}) =  \eta \mathbf{A}+ \mathbf{X}\boldsymbol{\beta},
\end{equation}
where $\mathbf{A}$ is the binary treatment assignment with $A=1$ indicating assignment to the intervention arm, $\eta$ is the treatment effect, and $\boldsymbol{\beta}$ is the vector including an intercept and covariate coefficients. The hypotheses are formulated as

\begin{equation*}
	H_0: \psi( \eta,\boldsymbol{\beta}) \leq \psi_0 \hskip5pt \text{vs} \hskip5pt H_A:  \psi(\eta, \boldsymbol{\beta})>\psi_0
\end{equation*}
where $\psi(\eta, \boldsymbol{\beta})$ is a marginal estimand such as the marginal risk ratio. Note that the direction of inequalities may need to be reversed depending on the definition of $\psi$. The Bayesian decision summary, $\tau$, and the decision rule is defined as in (\ref{eqn:pAlt}). The posterior distribution of $\psi$ cannot be obtained analytically using the logistic regression model in (\ref{eqn:adj}) since conjugate priors cannot be specified for the model parameters.  Following \cite{willard_covariate_2022}, the marginal estimand is defined as a contrast of the treatment group specific risk parameters. Then samples from the posterior of $\psi$ are obtained via Bayesian G-computation, marginalizing over the distribution of $\mathbf{X}$.

Consider a design exercise where frequentist operating characteristics and assurance are to be assessed for designs arising from various decision rules which are defined by values of $u$ and a range of (final or interim) sample sizes. In addition, within the covariate adjustment framework, comparison between various models ranging from the basic unadjusted model to a saturated model including all available covariates is of interest. 

In the following, we first focus on a simple model comparison exercise at the design stage and assess the proposed approach in estimating power and the type I error rate. Next, we consider estimating assurance in a hypothetical design exercise for a sub-study of a platform trial which seeks to study effectiveness of oral therapies against mild to moderate COVID-19 infection in individuals discharged from Canadian Emergency Departments. 

\subsection{Type I error rate and power curves for model comparison}
\label{sec:freqOC}
We use select simulation scenarios (Table~\ref{table1})  for dichotomous outcomes similar to those in \cite{willard_covariate_2022}. These scenarios are used to generate power and type I error rates as functions of $n$ and the effect parameters. The data generating model is the logistic regression model in (\ref{eqn:adj}) with $\mathbf{X} = (\mathbf{x}_1, \mathbf{x}_2, \mathbf{x}_3, \mathbf{x}_3^2, \mathbf{x}_5)$, generated from $F(\mathbf{X})$ (as specified in \cite{willard_covariate_2022}) and $\boldsymbol{\beta} = (-1.26, 1, -0.5, 1, -0.1, 0.5)$. For each of these scenarios, an adjusted model as described in (\ref{eqn:adj}) is compared with an unadjusted model,
\begin{equation}
	\label{eqn:unadj}
	{\rm logit} (\mathbf{p}) = \alpha \mathbf{1} + \gamma \mathbf{A}.
\end{equation}
where $\alpha$ is an intercept, $\mathbf{1}$ is a vector of ones, and $\gamma$ is the marginal treatment effect on the linear scale. The notation is selected to distinguish between the marginal ($\gamma$) and conditional ($\eta$) parameters.

We consider a simplified version of the simulation study of \cite{willard_covariate_2022}, and only compare the adjusted model with the ``correct" set of covariates to the unadjusted model. Additionally, we simulate a fixed design without stopping criteria at interim analyses so that we obtain the marginal sampling distribution of $\tau$ at a sequence of points throughout the trial. 

While this is a relatively simple setting for trial design simulation, considering $U$ values for $u$, $N$ values for sample size, $T$ values for $\eta$ and $\gamma$ (keeping other model parameters fixed), and $M$ simulation iterations (noting that $M>10,000$ is currently recommended by the \cite{fdacderbent_adaptive_nodate}) results in $\mathcal{O}(UNTM\mathcal{C})$, where $\mathcal{C}$ is the computational cost associated with posterior estimation. The proposed approach can reduce the computational complexity to $\mathcal{O}(N_0T_0M\mathcal{C})$, with $N_0<<N$ and $T_0<<T$.

\begin{figure}[h!]
	
	\begin{subfigure}[b]{0.47\textwidth}
		\centering
		\includegraphics[width=\textwidth]{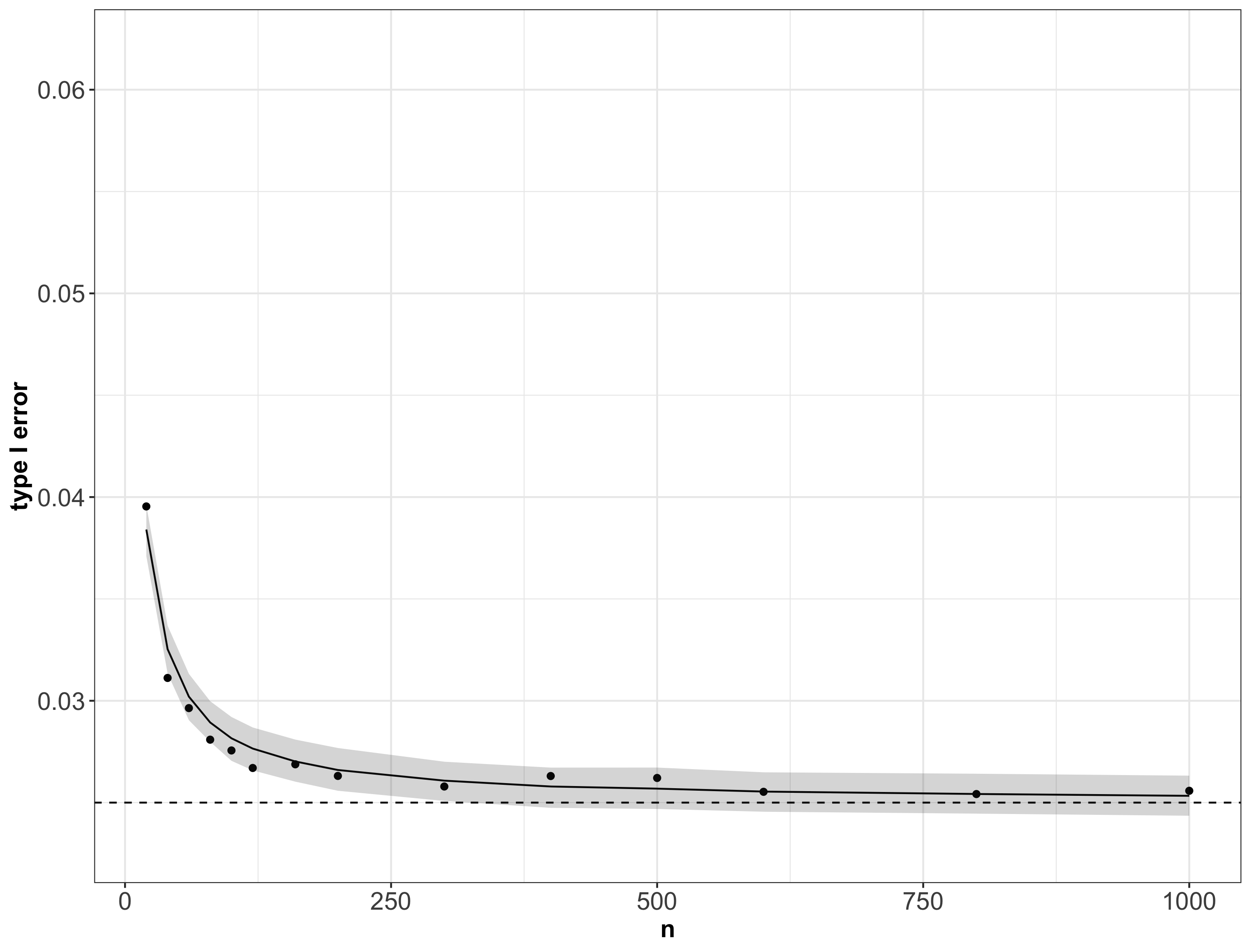}
		\caption{}
		\label{fig:t1eunadj}
	\end{subfigure}
	\begin{subfigure}[b]{0.47\textwidth}
		\centering
		\includegraphics[width=\textwidth]{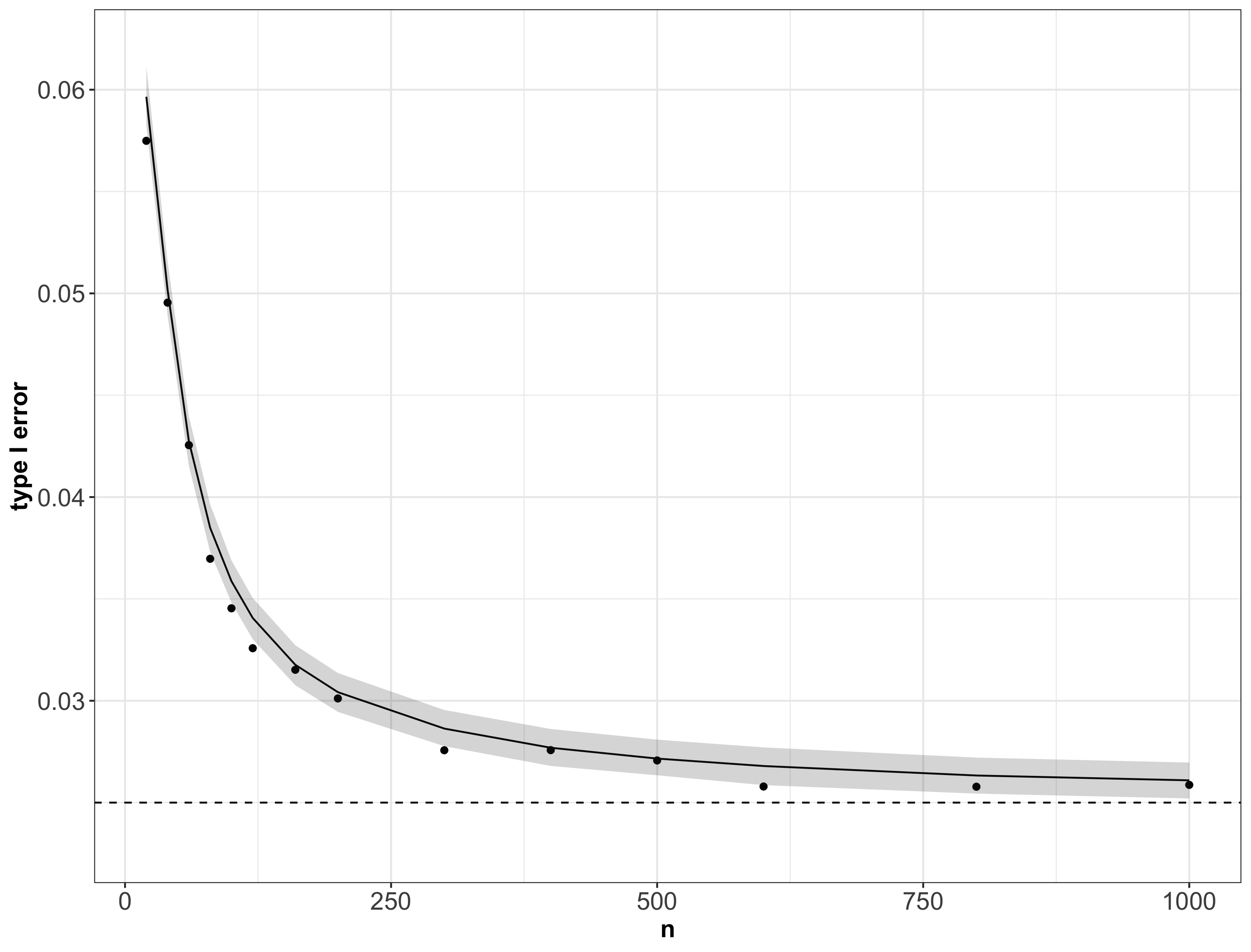}
		\caption{}
		\label{fig:t1eadj5}
	\end{subfigure}
	
	\caption{Estimated curves (posterior median) and equal-tailed point-wise 95\% credible intervals for type I error rate as a function of sample size in (a) unadjusted and (b) adjusted model. The dots are Monte Carlo based estimates of type I error rate.}
	\label{fig:t1e}
\end{figure}

 Figure~\ref{fig:t1e} shows the estimated curves (posterior median) and 95\% credible intervals (equal-tailed posterior quantiles) for type I error rate as a function of sample size for the adjusted and unadjusted models. These results are obtained by fitting the models in Section~\ref{Sec:methods} to the simulated sampling distribution (with $M = 100,000$ iterations) at the 14 sample sizes. The black solid circles are simulation-based estimates of the type I error rate. The goal of this exercise is to demonstrate that although these curves are generated by fitting a model to the sampling distribution and not directly to the simulated tail probabilities, they are able to capture the type I error rates obtained from the simulations and produce corresponding uncertainty estimates. The full curves exhibit the slower rate of convergence to the nominal type I error rate in the highly parameterized  adjusted model, a useful detail for specifying sample sizes and scheduling interim analyses when covariate adjustment is considered.

\begin{figure}[h!]
	
	\begin{subfigure}[b]{0.47\textwidth}
		\centering
		\includegraphics[width=\textwidth]{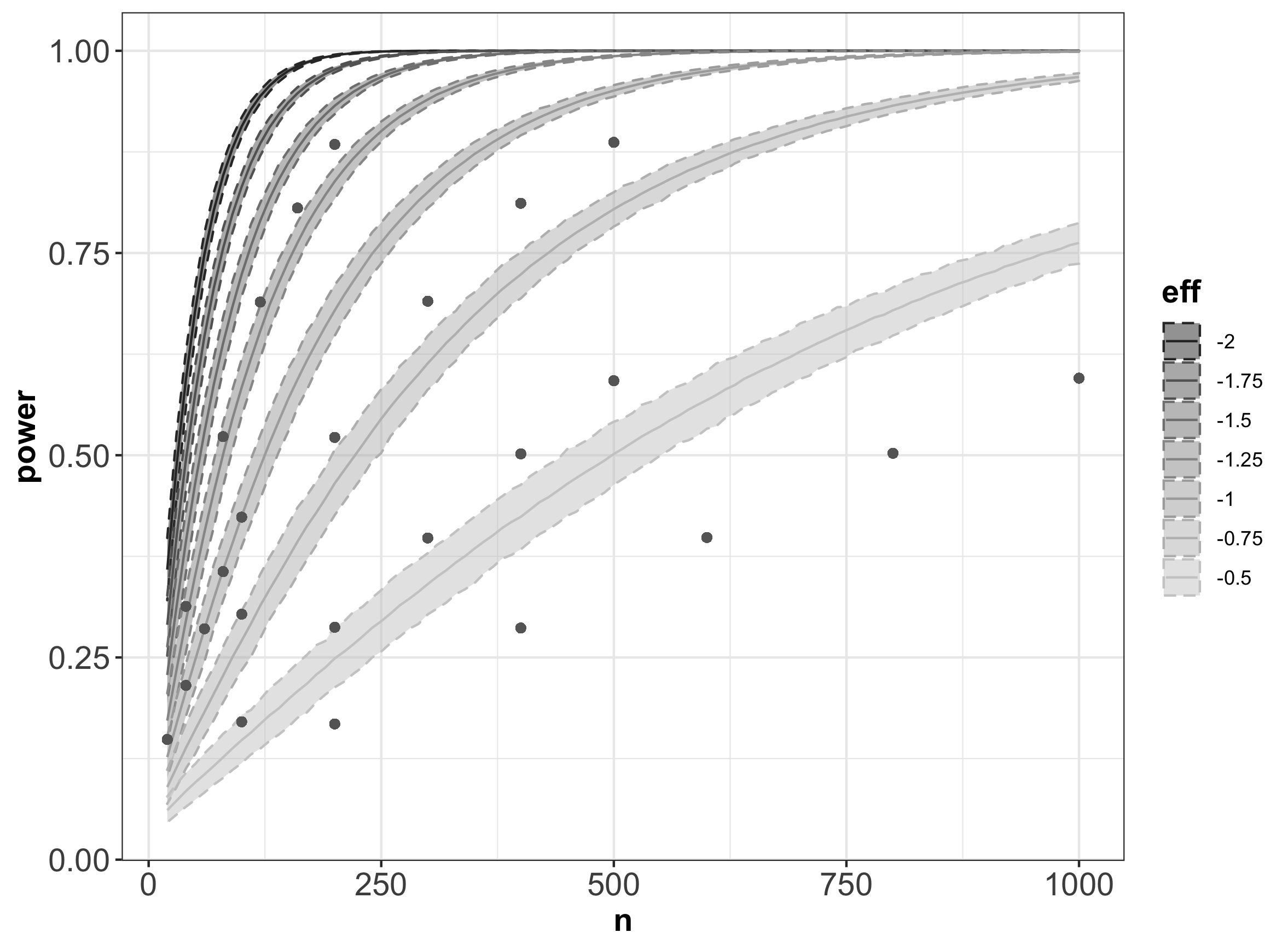}
		\caption{}
		\label{fig:power_all_unadj}
	\end{subfigure}
	\begin{subfigure}[b]{0.47\textwidth}
		\centering
		\includegraphics[width=\textwidth]{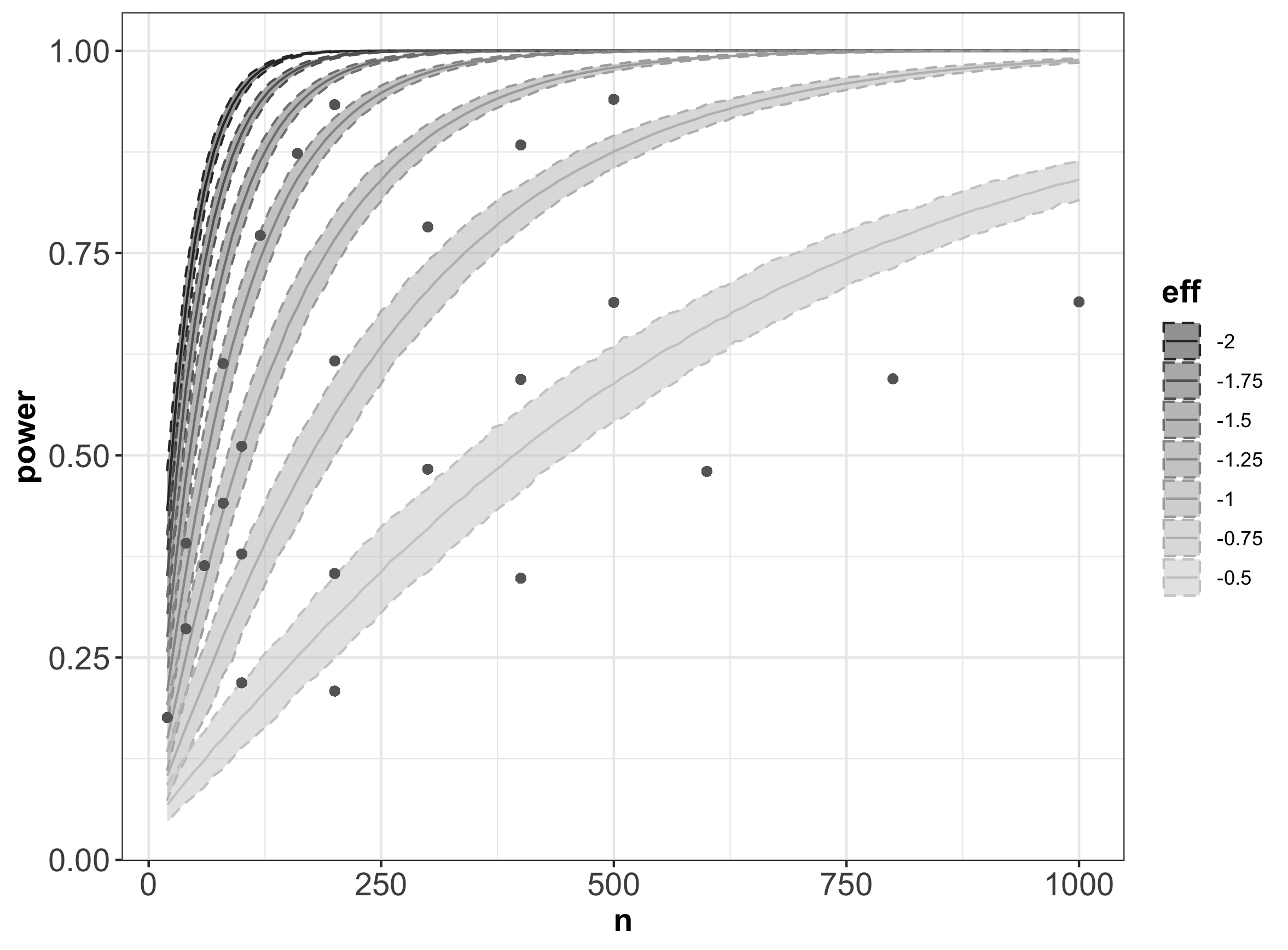}
		\caption{}
		\label{fig:power_all_adj}
	\end{subfigure}
	
	\caption{Estimated curves (posterior median) and equal-tailed point-wise 95\% credible intervals for power as a function of sample size for a range of effect sizes in the (a) unadjusted and (b) adjusted model. The dots are simulation-based estimates of power at selected points.}
	\label{fig:power}
\end{figure}

Figure~\ref{fig:power} shows the estimated curves and 95\% credible intervals for power as a function of sample size and a range of effect sizes which are different from those included in the simulation scenarios. The black solid circles are simulation-based estimates of power for the 25 scenarios listed in rows 2-6 of Table~\ref{table1}. This is to demonstrate that power curves and their associated uncertainty may be estimated for any given effect assumption of interest by modelling the sampling distribution using only 25 simulation scenarios. These estimates provide an overall understanding of power corresponding to each analysis and can be used to specify a design with the preferred analysis model.


The presented results showcase the use of the proposed approach but do not assess the performance compared to simulation-based estimates. To do so, we divide the simulation scenarios into a training set and a test set. The training set is used to fit the models proposed for the sampling distribution, which are then used to estimate the operating characteristics in the test set and compare them with those obtained from the simulations. 

\begin{figure}[h!]
	\centering
	\begin{subfigure}[b]{.75\textwidth}
		
		\includegraphics[width=\textwidth]{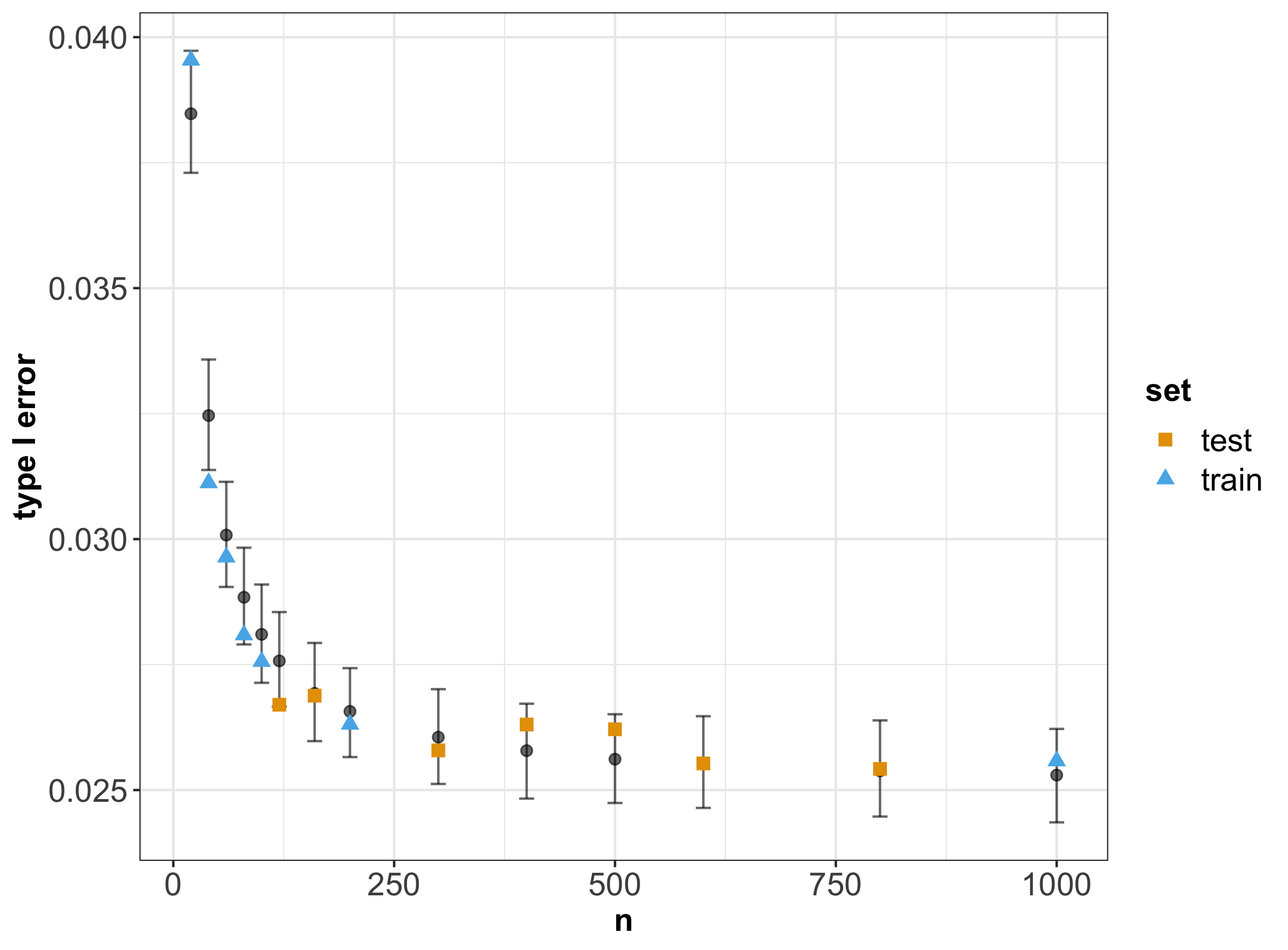}
		\caption{Unadjusted analysis model}
		\label{fig:test_unadj}
	\end{subfigure}\\
	\begin{subfigure}[b]{0.75\textwidth}
		
		\includegraphics[width=\textwidth]{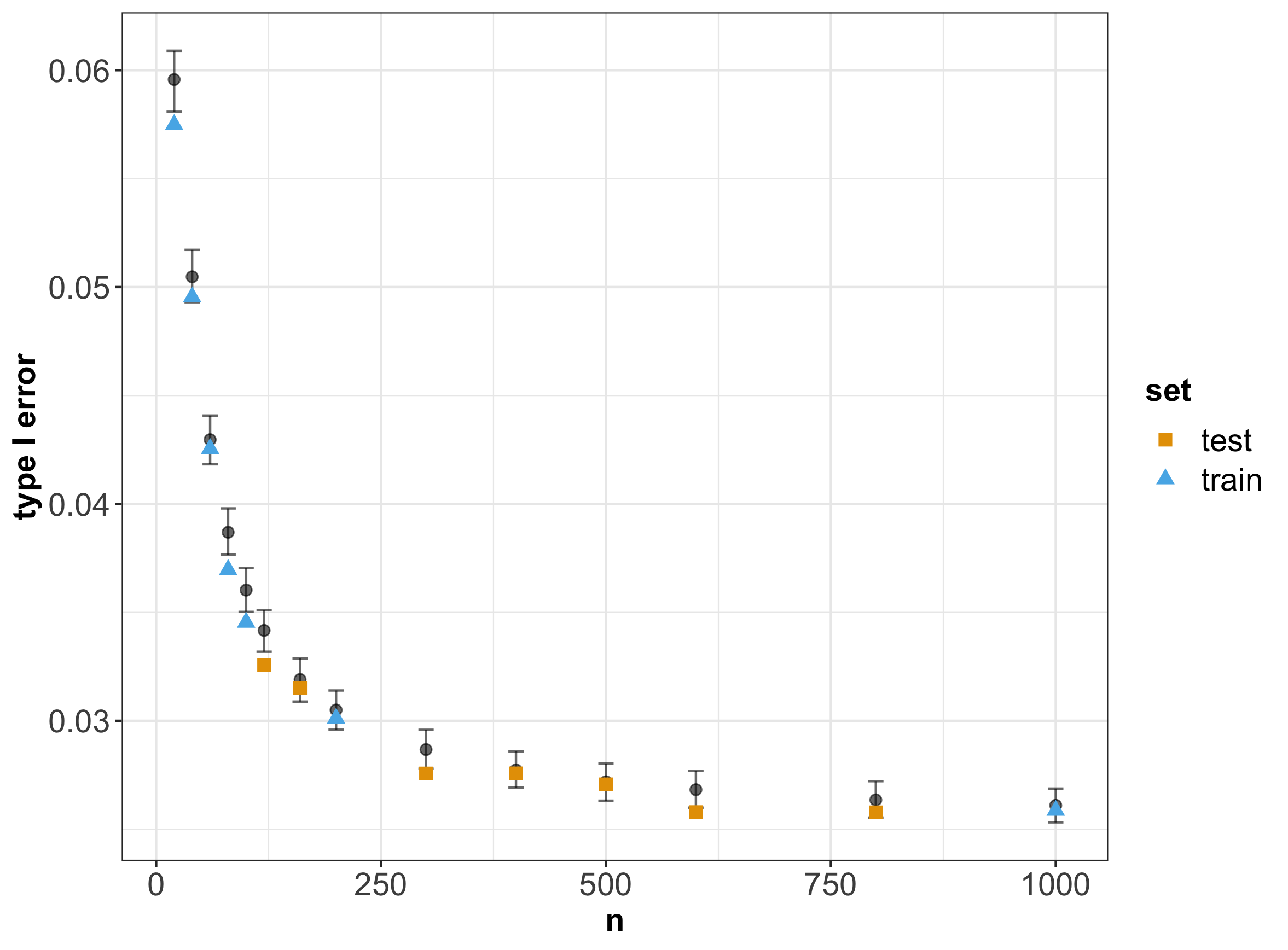}
		\caption{Adjusted analysis model}
		\label{fig:test_adj}
	\end{subfigure}
	
	\caption{Posterior median and equal-tailed 95\% credible intervals for type I error rate obtained from a training set of size 7 indicated by triangles. The squares represent the simulation based estimate of the type I error rate to be compared with the simulation-assisted estimate (dots).}
	\label{fig:test}
\end{figure}

Figure~\ref{fig:test} shows these results for the type I error rate, where the training set includes sample sizes $n = 20, 40, 60, 80, 100, 200, 1000$; the test set is defined as the remaining sample sizes listed in the first row of Table~\ref{table1}. The grey dots and error bars show estimates and 95\% credible intervals obtained from modelling the sampling distribution, while the triangles and squares show the simulation-based type I error rates in the training and test sets, respectively. For both models, type I error rate is estimated with less than 0.002 bias and in most cases, the 95\% credible intervals include the simulation-based estimates. 

The concentration of small sample sizes in the training set is intentional, as the evolution of the sampling distribution (and, as a result, type I error rate) for small $n$ is best learned from simulation results, while for larger sample sizes, the sampling distribution is well predicted by theoretical asymptotic properties.

To assess the accuracy and precision for power, we use the full set of simulation scenarios in \cite{willard_covariate_2022} that includes 15 additional effect and sample size value combinations, 	$n = (20, 40, 60, 80, 100)$ with $\eta = -1.24$; 	$n = (40, 80, 120, 160, 200)$ with $\eta =  -0.88$; and $n= (200, 400, 600, 800, 1000) $ with $\eta =  -0.55$. This results in a total of 40 simulation scenarios, which is divided into a training set of size 12 given in Table~\ref{table2}, and a test set that includes the 28 remaining scenarios. 

\begin{figure}[h!]
	\centering
	\begin{subfigure}[b]{.45\textwidth}
		
		\includegraphics[width=\textwidth]{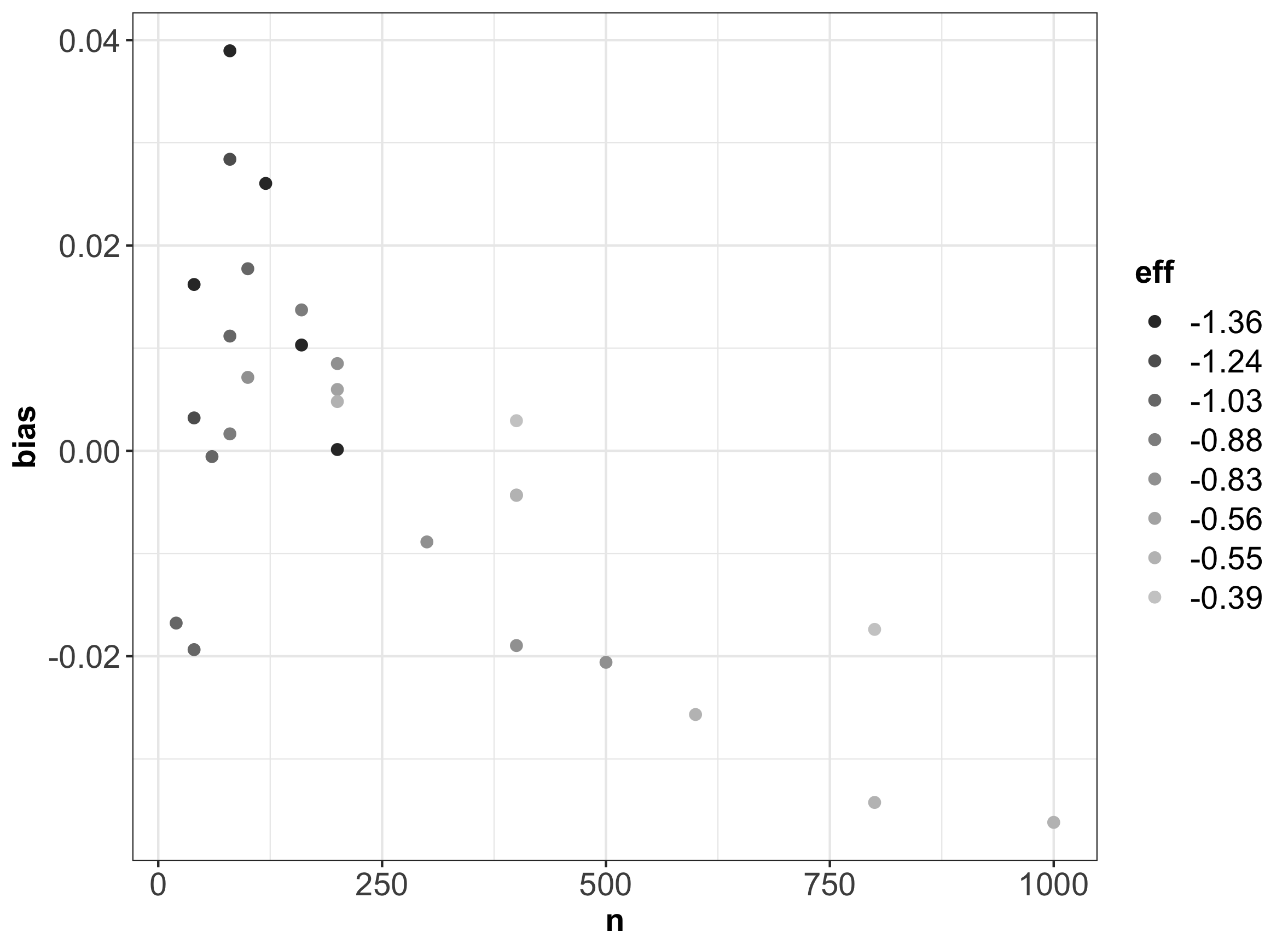}
		\caption{bias in the adjusted model}
		\label{fig:power_bias_adj}
	\end{subfigure}
	\begin{subfigure}[b]{.45\textwidth}
	
	\includegraphics[width=\textwidth]{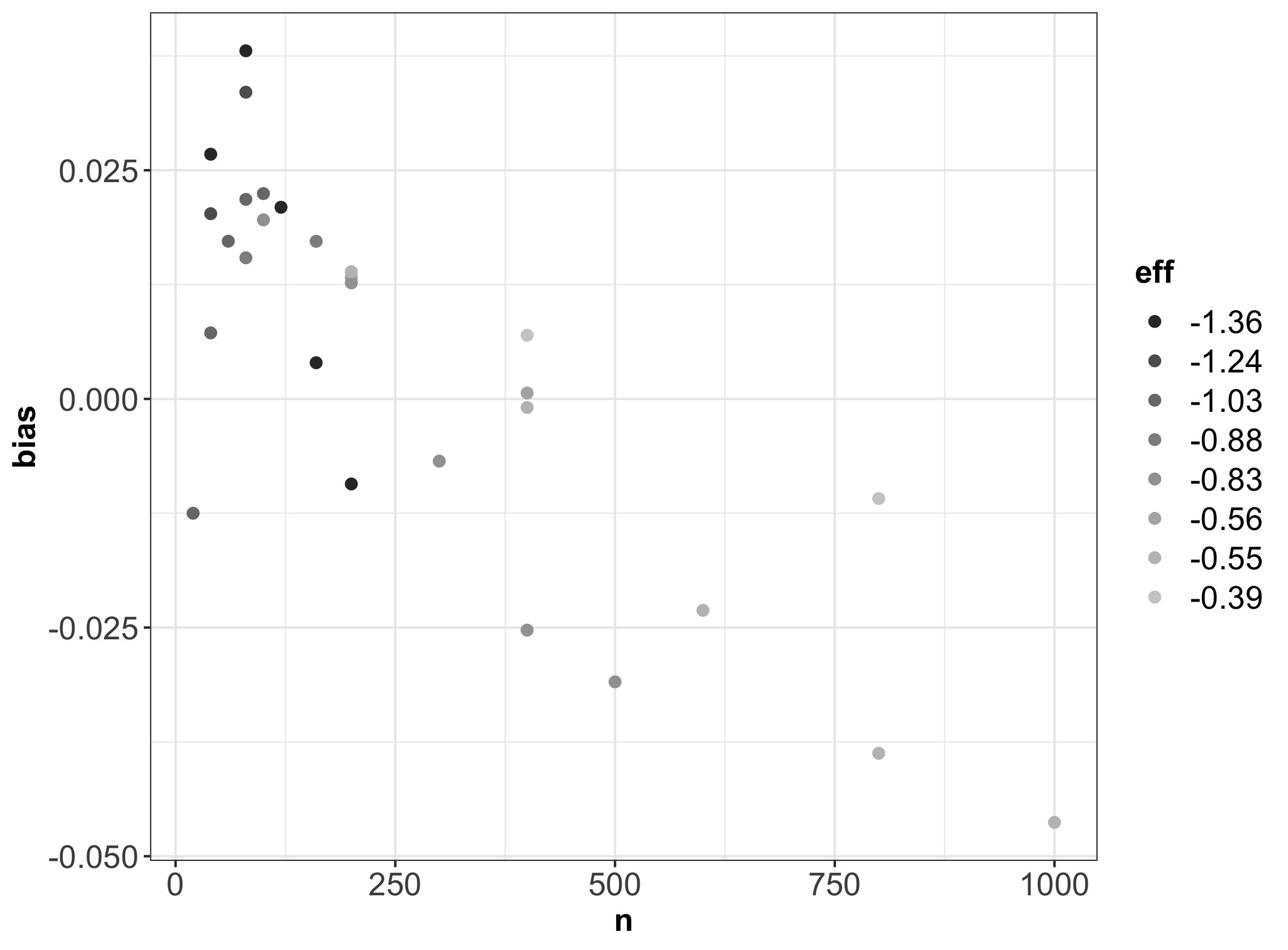}
	\caption{bias in the unadjusted model}
	\label{fig:power_bias_unadj}
\end{subfigure}\\
\begin{subfigure}[b]{0.45\textwidth}
	
	\includegraphics[width=\textwidth]{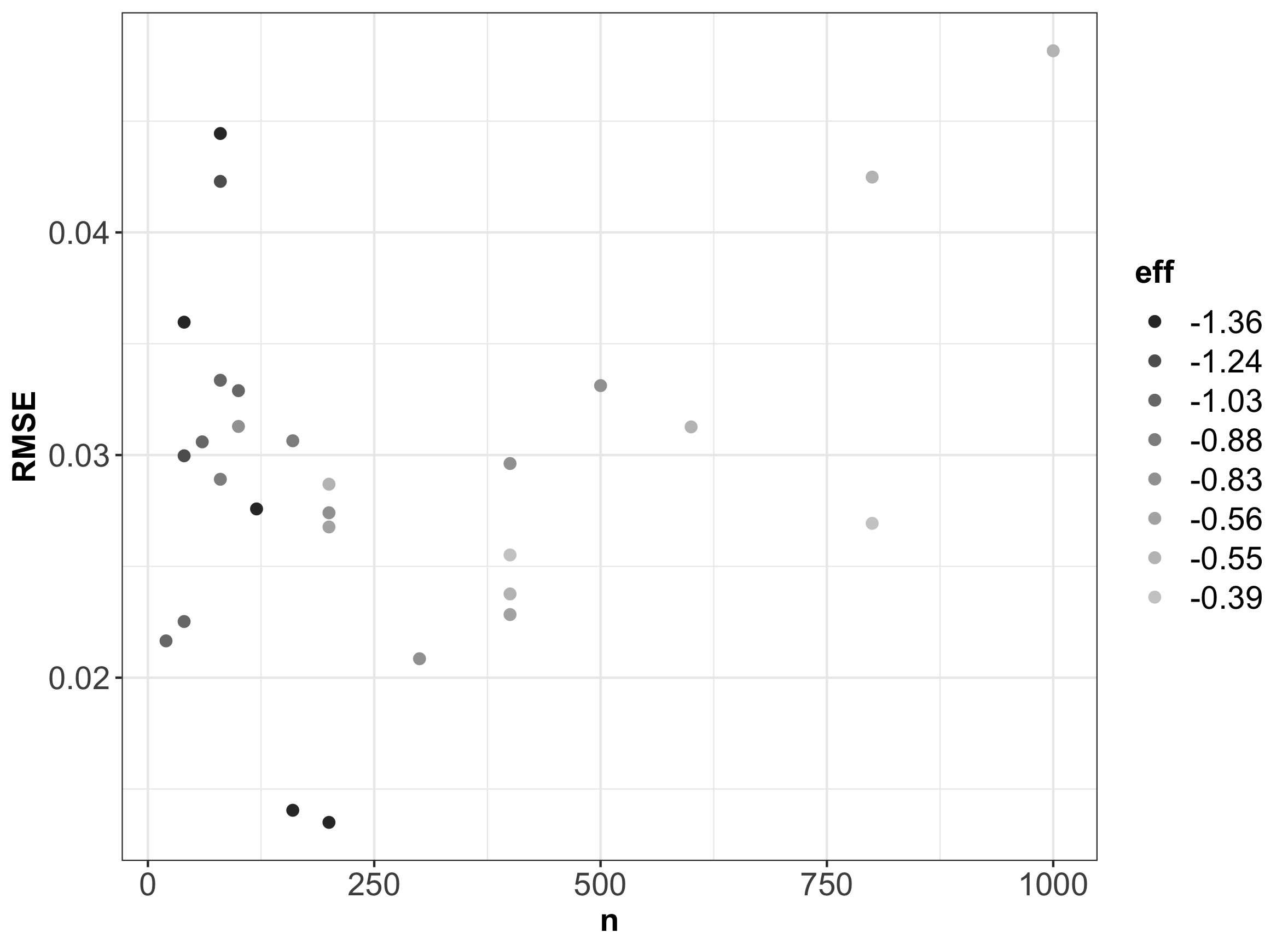}
	\caption{RMSE in the adjusted model}
	\label{fig:power_RMSE_unadj}
\end{subfigure}
	\begin{subfigure}[b]{0.45\textwidth}
		
		\includegraphics[width=\textwidth]{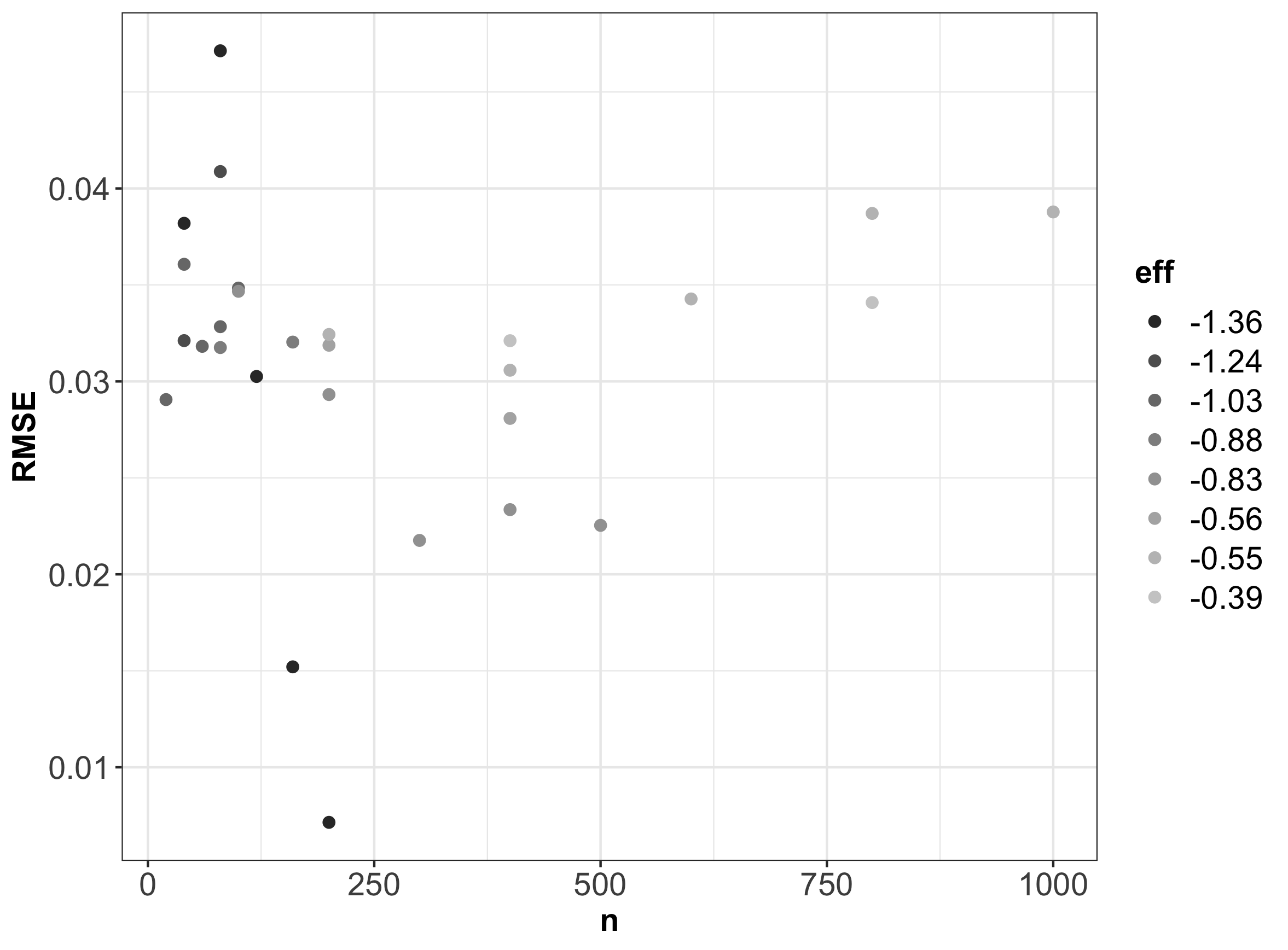}
		\caption{RMSE in the unadjusted model}
		\label{fig:power_RMSE_adj}
	\end{subfigure}
	
	\caption{Bias and RMSE in estimating power for a test set of size 28 from a training set of size 12 in the adjusted ((a) and (c))  and unadjusted ((b) and (d)) model.}
	\label{fig:test_power_adj}
\end{figure}

Instead of presenting graphs analogous to that of type I error rate, we plot the bias and root mean squared error (RMSE) for the 40 sample size/effect size scenarios in Figure~\ref{fig:test_power_adj} for both models. The results show that, across all scenarios and the two models, both the absolute bias and the RMSE remain below 0.05. Similar patterns are present for both models, where power is slightly overestimated for small to moderate $n$ and slightly underestimated for larger sample sizes. This bias, however, is inconsequential from a practical perspective.

\subsection{Bayesian assurance in an adaptive covariate-adjusted trial}
Having established precision and accuracy in the estimation of conditional operating characteristics, we now move on to demonstrate the capability of the proposed approach in a hypothetical design scenario where marginal operating characteristics (e.g., assurance) are of interest. We consider a sub-study of a platform design for evaluating the effectiveness of oral therapies against mild to moderate COVID-19 infection in individuals discharged from Canadian Emergency Departments. 

The trial design takes advantage of an already established network of physicians and researchers called the Canadian COVID-19 Emergency Department Rapid Response Network (CCEDRRN). We consider the sub-study where a single oral therapy is compared to the standard of care. The binary outcome of interest is a composite endpoint of 28-day hospitalization or mortality. Realistic values used in the trial simulation are taken from a COVID-19 Emergency Department risk model, developed by the CCEDRRN researchers \citep{hohl_ccedrrn_2022, hohl_treatments_2022, hohl_development_2021}. 

Suppose that there is a high degree of uncertainty associated with the hypothesized treatment effect. In particular, the trial investigators have differing prior views about the credible effect size and the uncertainty about it which can be expressed via two \emph{design} prior distributions over the conditional effect, $\eta$: an optimistic prior, $\mathcal{N}(-0.45, 0.01)$ centered at a relatively large effect size with a small level of uncertainty, and a conservative prior, $\mathcal{N}(-0.35, 0.05)$, centered at a smaller effect size with higher variability. The interest is to specify the first interim sample size such that the probability of establishing effectiveness over the design prior, i.e., assurance, is at least 20\% (an arbitrary threshold selected here for demonstration). 
		
\begin{figure}[h!]
	
	\centering
	\begin{subfigure}[b]{0.48\textwidth}
	
	\includegraphics[width=\textwidth]{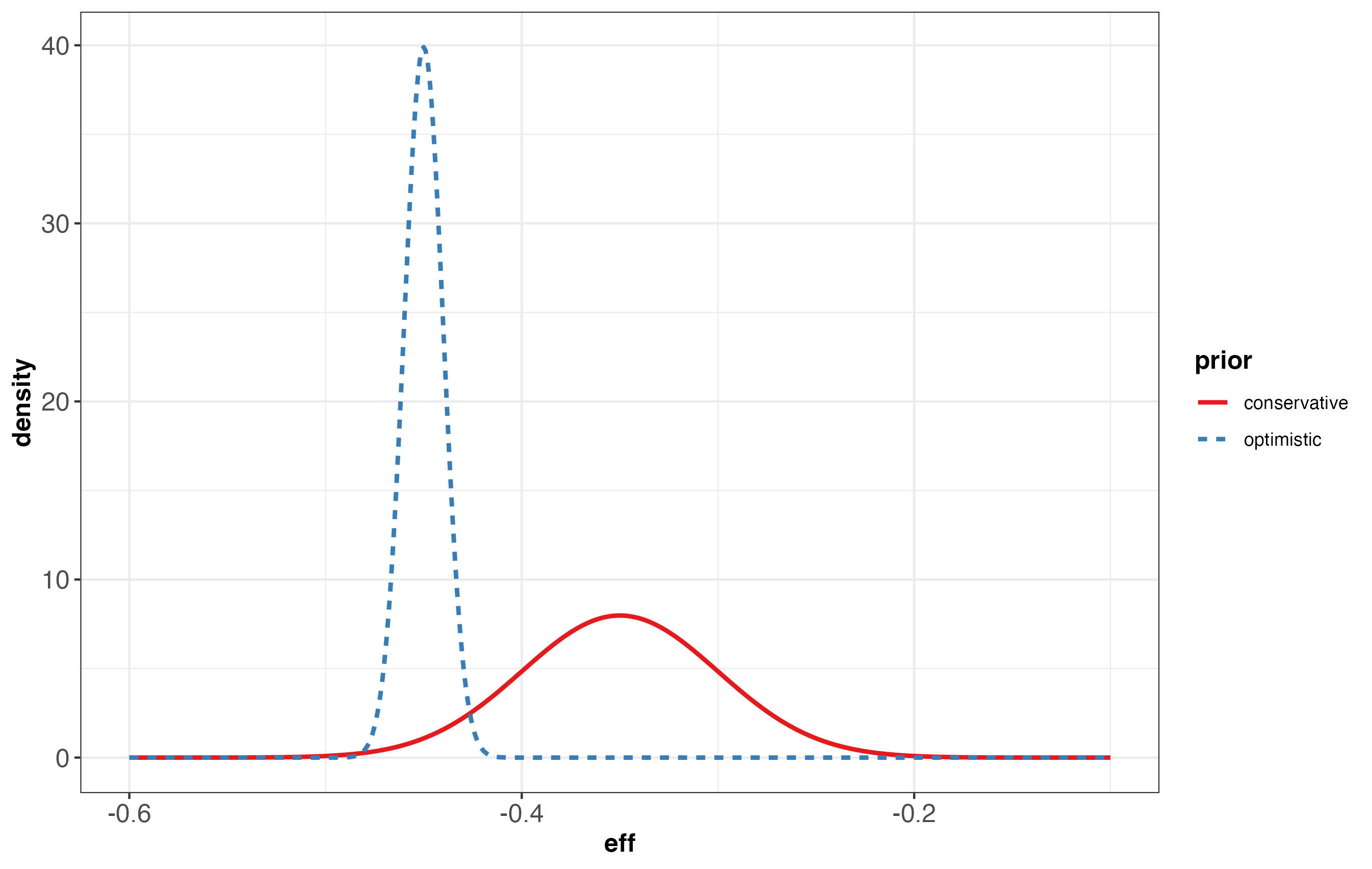}
	\caption{}
	\label{}
\end{subfigure}
	\begin{subfigure}[b]{0.48\textwidth}
		
		\includegraphics[width=\textwidth]{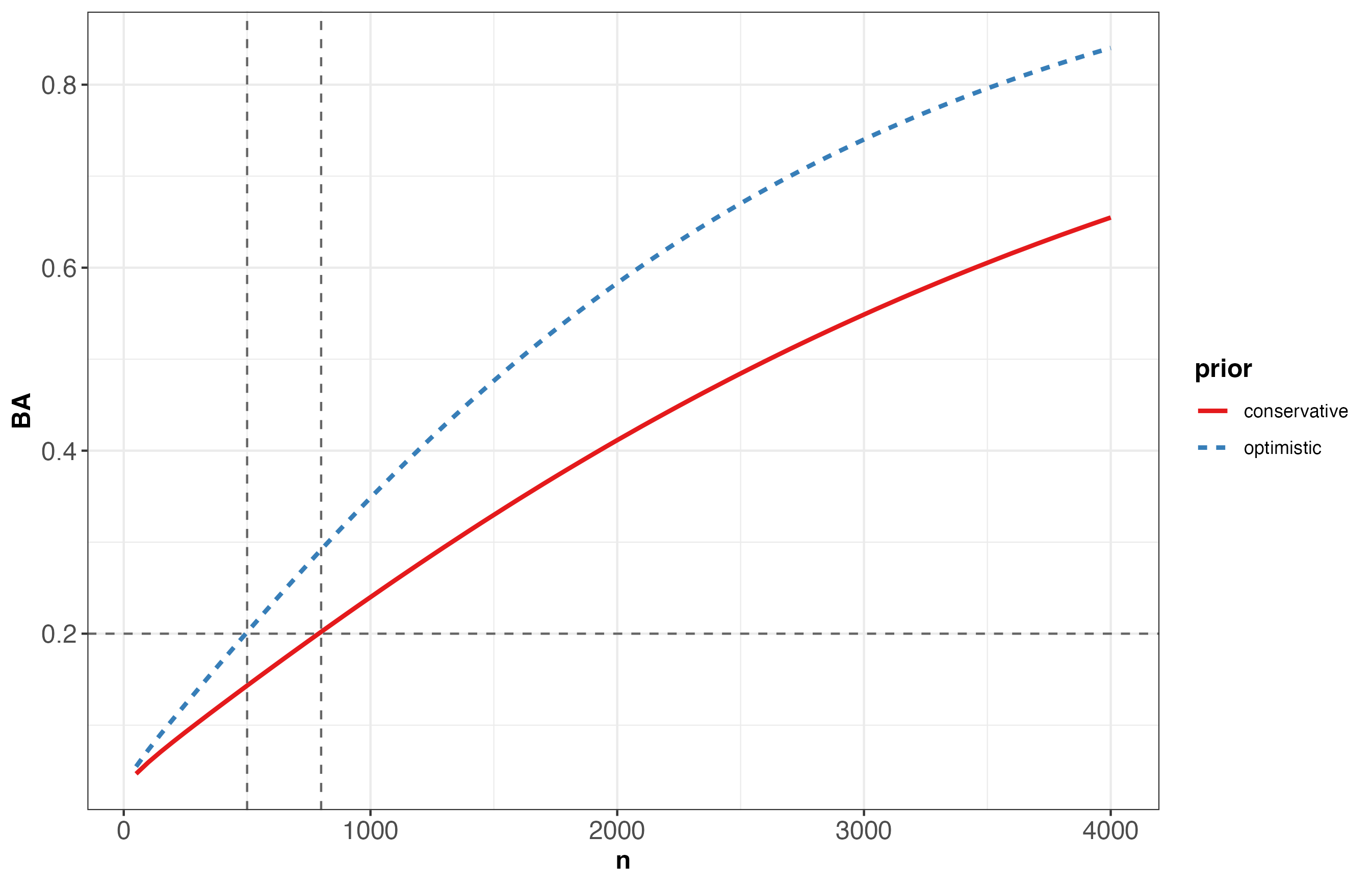}
		\caption{}
		\label{}
	\end{subfigure}
	
	\caption{(a) Two design priors representing different investigator views about the effect size and uncertainty associated with it; (b) Estimated Bayesian assurance (BA) curves as a function of sample size for the two design prior distribution. The dashed vertical lines are drawn at the interim sample size to achieve at least 20\% assurance under each scenario.}
	\label{fig:BA}
\end{figure}

To estimate  assurance with the proposed design priors, the sampling distribution is needed over the space which spans the two distributions' support and a reasonable range of sample sizes. An initial set of points over this space needs to be selected as a training set, i.e., the set of points where the sampling distribution is generated via Monte Carlo simulations. We use a coarse grid arising from the Cartesian product of effect sizes $\eta = (-0.6, -0.4, -0.2, -0.1)$ and sample sizes $n = (250, 500, 1000, 2000, 3000)$ resulting in 20 simulation scenarios. Power can then be estimated over the 2-d space and numerically integrated (using Monte Carlo) over the two proposed design priors resulting in the estimated Bayesian assurance which is plotted against the sample size in Figure~\ref{fig:BA}. The solid and dashed curves show Bayesian assurance for the conservative and optimistic priors, respectively, each of which uses a decision threshold of $u=0.975$. The graph suggests that to achieve at least 20\% assurance, the first interim analysis should be scheduled at $n = 500$ under the optimistic prior and $n= 800$ under the conservative prior. 
The nominal type I error rate was estimated below 3\% for the range of sample sizes considered in this design exercise. The assurance can be estimated for an alternative decision threshold at negligible additional cost, which is 1.5 minutes on a laptop computer with an Apple M2 (Pro) chip.

\subsection{Handling of nuisance parameters}

Note that in the presented examples, we have been treating the values of the nuisance parameters (covariate coefficients $\boldsymbol{\beta}$) as known. While in practice information about these parameters may be available, incorporating the uncertainty about these parameters in addition to the effect parameter is of interest. Consider the parametrization $\boldsymbol{\theta}= (\eta, \boldsymbol{\beta})$ and the design prior $\pi_d(\boldsymbol{\theta})= \pi_d(\eta)\pi(\boldsymbol{\beta})$. The Bayesian assurance can then be written as
\begin{equation}
	\text{BA} = \text{E}_{\boldsymbol{\theta}} \left[P\left(\tau>u\mid\boldsymbol{\theta} \right)\right]=   \int \left\{\int P\left(\tau(\mathbf{y})>u\mid \eta, \boldsymbol{\beta}\right)\pi_d(\eta)d\eta\right\} \pi_d(\boldsymbol{\beta}) d\boldsymbol{\beta}.\label{BA2}
\end{equation}
The proposed approach provides the inner integral by modeling the distribution of $\tau$ indexed by $\eta$. The outer integral can then be estimated numerically by repeating the described procedures over a grid for $\boldsymbol{\beta}$. An extension of the approach that models the sampling distribution over the extended parameter space can reduce the computational burden significantly. We discuss this extension in the next section as a direction of future research.

\section{Discussion}
\label{Sec:dis}
In this article, we propose a simulation-assisted approach for the assessment of conditional and marginal operating characteristics in Bayesian clinical trials that employ the posterior probability that the alternative hypothesis is true to make decisions. The proposed approach relies on modelling the sampling distribution of this posterior probability. Parametric models are specified to capture theoretical large sample properties of the posterior distribution and the expected behaviour of the decision procedure. The model parameters are estimated using ``observed" instances of the sampling distribution obtained via Monte Carlo simulations at select simulation scenarios. 

Theoretical results for the sampling distribution require large $n$, as they rely on asymptotic theory. However, operating characteristics for small to moderate sample sizes are of interest in complex designs. Therefore, existing methods for Bayesian sample size determination and assessment of design operating characteristics in clinical trials have mainly relied on Monte Carlo simulations. The novelty of the proposed approach is in utilizing asymptotic properties together with Monte Carlo simulations to learn the operating characteristics across a range of model/design parameters. This results in major computational savings for estimating conditional operating characteristics and enables estimation of marginal operating characteristics.

In principle, we agree with the argument that conditional operating characteristics are inadequate to assess clinical trials designs, especially under the Bayesian framework, since they ignore the uncertainty about the parameters at the design stage. This work is primarily intended to advocate for the use of marginal operating characteristics such as assurance. However, we realise that conditional operating characteristics continue to be required by regulatory agencies and used by many investigators. Therefore, providing efficient approaches to evaluate and report these criteria for Bayesian trials will enable wider use of Bayesian methods for clinical trials in practice.


While the present article is focused on the posterior probability that the alternative hypothesis is true, it can be modified to accommodate any other summary that is derived from the marginal posterior distribution of the parameter of interest in Bayesian trials. Moreover, similar methods may be developed to learn the sampling distribution of test statistics in non-Bayesian clinical trials.

The selection of simulation scenarios at which the sampling distribution is simulated is important, as these simulations play the role of \emph{data} in the proposed approach. The selection of these scenarios is therefore a \emph{design} problem. For type I error rate, this boils down to selecting a sequence of sample size values. We recommend concentrating the simulations at small $n$, where the type I error rate changes (decreases) quickly with sample size. At large $n$, the probability of a type I error converges to $(1-u)\%$, where $u$ is the decision threshold, and the additional refinement from simulations is negligible.  A similar rationale applies in determining the simulation scenarios for estimating power curves. An equally-spaced sequence over the support of the design prior paired with a sequence of reasonable sample sizes achieves acceptable performance.

The methods proposed in this article are advantageous in a variety of complex design and analysis settings. For example, \cite{chapple_multi-armed_2023} proposed a multi-arm Bayesian utility-based sequential trial design with a pairwise
null clustering prior where their simulation study involved sampling parameters from a (design) distribution. Given the complexity of the design, statistical model and simulation study, this is an example of a case where the proposed methods in this manuscript would apply. Other settings where this methodology can make a notable difference include clinical trials with Bayesian hierarchical models (e.g., \cite{liu_bayesian_2020}, \cite{carragher_bayesian_2020}, and \cite{zhao_bayesian_nodate}), trials that employ information borrowing techniques resulting in analytically intractable posterior distributions (for example, \cite{psioda_practical_2018} and \cite{zhou_incorporating_2021}), and trials that incorporate time trends in the analysis (e.g., \cite{saville_bayesian_2022}). In addition, the proposed approach can facilitate decision-theoretic approaches toward clinical trial design (e.g., \cite{lewis_bayesian_2007} and \cite{calderazzo_decision-theoretic_2020}). 

We emphasize that the proposed approach can be employed regardless of the computational method used to obtain the posterior distribution and its summaries. In our examples, we rely on MCMC sampling for Bayesian analysis. However, other sampling or approximation methods such as Integrated Nested Laplace Approximation \citep{rue_approximate_2009, hosseini_designing_2023} or variational methods \citep{tran_variational_2022} can be used to further reduce the computational cost in the first stage.

The proposed methodology can be extended in various directions. In this article, we focus on the marginal distribution of a posterior summary at interim analyses. To increase the applicability of the proposed approach to adaptive designs, the joint sampling distribution over multiple interim analyses could be modeled. Also, in this work, we perform power analyses that explore various settings of the effect parameter but hold nuisance parameters fixed. While this reflects common practice in trial design, being able to handle uncertainty with respect to the nuisance parameters is a great advantage of Bayesian designs. Therefore, an important extension is to model the sampling distribution across the extended parameter space. 

To conclude, we emphasize that the proposed approach is one that facilitates and complements various existing methods in the Bayesian SSD and trial design literature. We anticipate that this work will greatly contribute to the practice of statistical trial design for trials with complex analysis models and innovative/flexible designs.

\section*{Acknowledgements}
The authors would like to thank Dr Richard Lockhart and Dr Alexandra Schmidt as well as the peer-review team for their comments on the manuscript. In addition, we thank Dr Corinne Hohl for the information on the CCEDRRN example.

SG acknowledges support from the Natural Sciences and Engineering Research Council of Canada (NSERC), Candian Institute for Statistical Sciences (CANSSI) and Fonds de recherche du Québec - Santé (FRQS). JW acknowledges the support of a doctoral training scholarship from the Fonds de recherche du Québec - Nature et technologies (FRQNT).

\bibliographystyle{apalike}
\bibliography{ref02}

\begin{thebibliography}{}

\bibitem[Benkeser et~al., 2021]{benkeser2021improving}
Benkeser, D., D{\'\i}az, I., Luedtke, A., Segal, J., Scharfstein, D., and
  Rosenblum, M. (2021).
\newblock Improving precision and power in randomized trials for {COVID-19}
  treatments using covariate adjustment, for binary, ordinal, and time-to-event
  outcomes.
\newblock {\em Biometrics}, 77(4):1467--1481.

\bibitem[Berry, 1993]{berry_case_1993}
Berry, D.~A. (1993).
\newblock A case for {Bayesianism} in clinical trials.
\newblock {\em Statistics in Medicine}, 12(15-16):1377--1393; discussion
  1395--1404.

\bibitem[Berry et~al., 2010]{berry_bayesian_nodate}
Berry, S.~M., Carlin, B.~P., Lee, J.~J., and Muller, P. (2010).
\newblock Bayesian adaptive methods for clinical trials.
\newblock page 316.

\bibitem[Bissiri et~al., 2016]{bissiri_general_2016}
Bissiri, P.~G., Holmes, C.~C., and Walker, S.~G. (2016).
\newblock A general framework for updating belief distributions.
\newblock {\em Journal of the Royal Statistical Society. Series B (Statistical
  Methodology)}, 78(5):1103--1130.
\newblock Publisher: [Royal Statistical Society, Wiley].

\bibitem[Calderazzo et~al., 2020]{calderazzo_decision-theoretic_2020}
Calderazzo, S., Wiesenfarth, M., and Kopp-Schneider, A. (2020).
\newblock A decision-theoretic approach to {Bayesian} clinical trial design and
  evaluation of robustness to prior-data conflict.
\newblock {\em Biostatistics (Oxford, England)}, 23(1):328--344.

\bibitem[Carragher et~al., 2020]{carragher_bayesian_2020}
Carragher, R., Mueller, T., Bennie, M., and Robertson, C. (2020).
\newblock A {Bayesian} hierarchical approach for multiple outcomes in routinely
  collected healthcare data.
\newblock {\em Statistics in Medicine}, 39(20):2639--2654.
\newblock \_eprint: https://onlinelibrary.wiley.com/doi/pdf/10.1002/sim.8563.

\bibitem[Chapple et~al., 2023]{chapple_multi-armed_2023}
Chapple, A., Bennani, Y., and Clement, M. (2023).
\newblock A multi-armed bayesian ordinal outcome utility-based sequential trial
  with a pairwise null clustering prior.
\newblock {\em Bayesian Analysis}, 18(2):519--546.
\newblock Publisher: International Society for Bayesian Analysis.

\bibitem[Chuang-Stein, 2006]{chuang-stein_sample_2006}
Chuang-Stein, C. (2006).
\newblock Sample size and the probability of a successful trial.
\newblock {\em Pharmaceutical Statistics}, 5(4):305--309.
\newblock \_eprint: https://onlinelibrary.wiley.com/doi/pdf/10.1002/pst.232.

\bibitem[FDA, 2019]{fdacderbent_adaptive_nodate}
FDA (2019).
\newblock Adaptive designs for clinical trials of drugs and biologics.
\newblock FDA-2018-D-3124.

\bibitem[Gelfand and Wang, 2002]{gelfand_simulation-based_2002}
Gelfand, A.~E. and Wang, F. (2002).
\newblock A simulation-based approach to {Bayesian} sample size determination
  for performance under a given model and for separating models.
\newblock {\em Statistical Science}, 17(2):193--208.
\newblock Publisher: Institute of Mathematical Statistics.

\bibitem[Golchi, 2022]{golchi_estimating_2022}
Golchi, S. (2022).
\newblock Estimating design operating characteristics in {Bayesian} adaptive
  clinical trials.
\newblock {\em Canadian Journal of Statistics}, 50(2):417--436.
\newblock \_eprint: https://onlinelibrary.wiley.com/doi/pdf/10.1002/cjs.11699.

\bibitem[Hampson et~al., 2022]{HamBorHol22}
Hampson, L.~V., Bornkamp, B., Holzhauer, B., Kahn, J., Lange, M.~R., Luo,
  W.-L., Cioppa, G.~D., Stott, K., and Ballerstedt, S. (2022).
\newblock Improving the assessment of the probability of success in late stage
  drug development.
\newblock {\em Pharmaceutical Statistics}, 21(2):439--459.

\bibitem[Han et~al., 2024]{HAT_2024}
Han, L., Arfè, A., and Trippa, L. (2024).
\newblock Sensitivity analyses of clinical trial designs: Selecting scenarios
  and summarizing operating characteristics.
\newblock {\em The American Statistician}, 78(1).

\bibitem[Hohl et~al., 2022a]{hohl_ccedrrn_2022}
Hohl, C.~M., Rosychuk, R.~J., Archambault, P.~M., O'Sullivan, F., Leeies, M.,
  Mercier, E., Clark, G., Innes, G.~D., Brooks, S.~C., Hayward, J., Ho, V.,
  Jelic, T., Welsford, M., Sivilotti, M. L.~A., Morrison, L.~J., Perry, J.~J.,
  and {Canadian COVID-19 Emergency Department Rapid Response Network (CCEDRRN)
  investigators for the Network of Canadian Emergency Researchers and the
  Canadian Critical Care Trials Group} (2022a).
\newblock The {CCEDRRN} {COVID}-19 mortality score to predict death among
  nonpalliative patients with {COVID}-19 presenting to emergency departments: a
  derivation and validation study.
\newblock {\em {CMAJ} open}, 10(1):E90--E99.

\bibitem[Hohl et~al., 2022b]{hohl_treatments_2022}
Hohl, C.~M., Rosychuk, R.~J., Hau, J.~P., Hayward, J., Landes, M., Yan, J.~W.,
  Ting, D.~K., Welsford, M., Archambault, P.~M., Mercier, E., Chandra, K.,
  Davis, P., Vaillancourt, S., Leeies, M., Small, S., Morrison, L.~J., and
  {Canadian COVID-19 Rapid Response Network (CCEDRRN) investigators for the
  Network of Canadian Emergency Researchers, for the Canadian Critical Care
  Trials Group} (2022b).
\newblock Treatments, resource utilization, and outcomes of {COVID}-19 patients
  presenting to emergency departments across pandemic waves: an observational
  study by the canadian {COVID}-19 emergency department rapid response network
  ({CCEDRRN}).
\newblock {\em {CJEM}}, 24(4):397--407.

\bibitem[Hohl et~al., 2021]{hohl_development_2021}
Hohl, C.~M., Rosychuk, R.~J., McRae, A.~D., Brooks, S.~C., Archambault, P.,
  Fok, P.~T., Davis, P., Jelic, T., Turner, J.~P., Rowe, B.~H., Mercier, E.,
  Cheng, I., Taylor, J., Daoust, R., Ohle, R., Andolfatto, G., Atzema, C.,
  Hayward, J., Khangura, J.~K., Landes, M., Lang, E., Martin, I., Mohindra, R.,
  Ting, D.~K., Vaillancourt, S., Welsford, M., Brar, B., Dahn, T., Wiemer, H.,
  Yadav, K., Yan, J.~W., Stachura, M., McGavin, C., Perry, J.~J., Morrison,
  L.~J., and {Canadian COVID-19 Emergency Department Rapid Response Network
  investigators and for the Network of Canadian Emergency Researchers and the
  Canadian Critical Care Trials Group} (2021).
\newblock Development of the canadian {COVID}-19 emergency department rapid
  response network population-based registry: a methodology study.
\newblock 9(1):E261--E270.

\bibitem[Hosseini et~al., 2023]{hosseini_designing_2023}
Hosseini, R., Chen, Z., Goligher, E., Fan, E., Ferguson, N.~D., Harhay, M.~O.,
  Sahetya, S., Urner, M., Yarnell, C.~J., and Heath, A. (2023).
\newblock Designing a {Bayesian} adaptive clinical trial to evaluate novel
  mechanical ventilation strategies in acute respiratory failure using
  {Integrated} {Nested} {Laplace} {Approximations}.
\newblock arXiv:2303.18093 [stat].

\bibitem[Kunzmann et~al., 2021]{kunzmann_review_2021}
Kunzmann, K., Grayling, M.~J., Lee, K.~M., Robertson, D.~S., Rufibach, K., and
  Wason, J. M.~S. (2021).
\newblock A review of {Bayesian} perspectives on sample size derivation for
  confirmatory trials.
\newblock {\em The American Statistician}, 75(4):424--432.

\bibitem[Lee et~al., 2022]{lee2022benefits}
Lee, K.~M., Robertson, D.~S., Jaki, T., and Emsley, R. (2022).
\newblock {The Benefits of Covariate Adjustment for Adaptive Multi-arm
  Designs}.
\newblock {\em Statistical Methods in Medical Research}, 31(11):2104--2121.

\bibitem[Lewis et~al., 2007]{lewis_bayesian_2007}
Lewis, R.~J., Lipsky, A.~M., and Berry, D.~A. (2007).
\newblock Bayesian decision-theoretic group sequential clinical trial design
  based on a quadratic loss function: a frequentist evaluation.
\newblock {\em Clinical Trials}, 4(1):5--14.
\newblock Publisher: SAGE Publications.

\bibitem[Liu et~al., 2020]{liu_bayesian_2020}
Liu, J., Wick, J., Jiang, Y., Mayo, M., and Gajewski, B. (2020).
\newblock Bayesian {Accrual} {Modeling} and {Prediction} in {Multicenter}
  {Clinical} {Trials} with {Varying} {Center} {Activation} {Times}.
\newblock {\em Pharmaceutical statistics}, 19(5):692--709.

\bibitem[Müller, 2005]{muller_simulation_2005}
Müller, P. (2005).
\newblock Simulation based optimal design.
\newblock In {\em Handbook of {Statistics}}, volume~25, pages 509--518.
  Elsevier.
\newblock Edited by D.K. Dey, C.R. Rao.

\bibitem[Müller and Parmigiani, 1995]{muller_optimal_1995}
Müller, P. and Parmigiani, G. (1995).
\newblock Optimal design via curve fitting of {Monte} {Carlo} experiments.
\newblock {\em Journal of the American Statistical Association},
  90(432):1322--1330.
\newblock Publisher: [American Statistical Association, Taylor \& Francis,
  Ltd.].

\bibitem[O'Hagan et~al., 2005]{ohagan_assurance_2005}
O'Hagan, A., Stevens, J.~W., and Campbell, M.~J. (2005).
\newblock Assurance in clinical trial design.
\newblock {\em Pharmaceutical Statistics}, 4(3):187--201.
\newblock \_eprint: https://onlinelibrary.wiley.com/doi/pdf/10.1002/pst.175.

\bibitem[O’Hagan and Stevens, 2001]{ohagan_bayesian_2001}
O’Hagan, A. and Stevens, J.~W. (2001).
\newblock Bayesian assessment of sample size for clinical trials of
  cost-effectiveness.
\newblock {\em Medical Decision Making}, 21(3):219--230.
\newblock Publisher: SAGE Publications Inc STM.

\bibitem[Pan and Banerjee, 2021]{pan_unifying_2021}
Pan, J. and Banerjee, S. (2021).
\newblock A unifying {Bayesian} approach for sample size determination using
  design and analysis priors.
\newblock arXiv:2112.03509 [stat].

\bibitem[Psioda et~al., 2018]{psioda_practical_2018}
Psioda, M.~A., Soukup, M., and Ibrahim, J.~G. (2018).
\newblock A practical {Bayesian} adaptive design incorporating data from
  historical controls.
\newblock {\em Statistics in medicine}, 37(27):4054--4070.

\bibitem[Ren and Oakley, 2014]{ren_assurance_2014}
Ren, S. and Oakley, J.~E. (2014).
\newblock Assurance calculations for planning clinical trials with
  time-to-event outcomes.
\newblock {\em Statistics in Medicine}, 33(1):31--45.

\bibitem[Rue et~al., 2009]{rue_approximate_2009}
Rue, H., Martino, S., and Chopin, N. (2009).
\newblock Approximate {Bayesian} inference for latent {Gaussian} models by
  using integrated nested {Laplace} approximations.
\newblock {\em Journal of the Royal Statistical Society: Series B (Statistical
  Methodology)}, 71(2):319--392.
\newblock \_eprint:
  https://onlinelibrary.wiley.com/doi/pdf/10.1111/j.1467-9868.2008.00700.x.

\bibitem[Saville et~al., 2022]{saville_bayesian_2022}
Saville, B.~R., Berry, D.~A., Berry, N.~S., Viele, K., and Berry, S.~M. (2022).
\newblock The {Bayesian} time machine: {Accounting} for temporal drift in
  multi-arm platform trials.
\newblock {\em Clinical Trials}, page 17407745221112013.
\newblock Publisher: SAGE Publications.

\bibitem[Sgouropoulos et~al., 2015]{sgouropoulos_matching_2015}
Sgouropoulos, N., Yao, Q., and Yastremiz, C. (2015).
\newblock Matching a distribution by matching quantiles estimation.
\newblock {\em Journal of the American Statistical Association},
  110(510):742--759.

\bibitem[Spiegelhalter et~al., 2003]{spiegelhalter_bayesian_2003}
Spiegelhalter, D.~J., Abrams, K.~R., and Myles, J.~P. (2003).
\newblock {\em Bayesian Approaches to Clinical Trials and Health-Care
  Evaluation}.
\newblock John Wiley \& Sons, Ltd, 1 edition.
\newblock \_eprint: https://onlinelibrary.wiley.com/doi/pdf/10.1002/0470092602.

\bibitem[Spiegelhalter and Freedman, 1986]{spiegelhalter_predictive_1986}
Spiegelhalter, D.~J. and Freedman, L.~S. (1986).
\newblock A predictive approach to selecting the size of a clinical trial,
  based on subjective clinical opinion.
\newblock {\em Statistics in Medicine}, 5(1):1--13.
\newblock \_eprint:
  https://onlinelibrary.wiley.com/doi/pdf/10.1002/sim.4780050103.

\bibitem[Tran et~al., 2022]{tran_variational_2022}
Tran, M.-N., Nott, D., and Kohn, R. (2022).
\newblock Variational {Bayes}.
\newblock In {\em Wiley {StatsRef}: {Statistics} {Reference} {Online}}, pages
  1--9. John Wiley \& Sons, Ltd.
\newblock \_eprint:
  https://onlinelibrary.wiley.com/doi/pdf/10.1002/9781118445112.stat08387.

\bibitem[{van der Vaart}, 1998]{vaart_asymptotic_1998}
{van der Vaart}, A.~W. (1998).
\newblock {\em Asymptotic {Statistics}}.
\newblock Cambridge {Series} in {Statistical} and {Probabilistic}
  {Mathematics}. Cambridge University Press, Cambridge.

\bibitem[{van Lancker} et~al., 2022]{van2022combining}
{van Lancker}, K., Betz, J., and Rosenblum, M. (2022).
\newblock Combining covariate adjustment with group sequential and information
  adaptive designs to improve randomized trial efficiency.
\newblock arXiv:2201.12921 [stat].

\bibitem[Willard et~al., 2022]{willard_covariate_2022}
Willard, J., Golchi, S., and Moodie, E.~E. (2022).
\newblock Covariate adjustment in {Bayesian} adaptive clinical trials.
\newblock arXiv:2212.08968 [stat].

\bibitem[Zhao et~al., 2022]{zhao_bayesian_nodate}
Zhao, Y., Tang, R.~S., Du, Y., and Yuan, Y. (2022).
\newblock A {Bayesian} platform trial design to simultaneously evaluate
  multiple drugs in multiple indications with mixed endpoints.
\newblock {\em Biometrics}.
\newblock \_eprint: https://onlinelibrary.wiley.com/doi/pdf/10.1111/biom.13694.

\bibitem[Zhou and Ji, 2021]{zhou_incorporating_2021}
Zhou, T. and Ji, Y. (2021).
\newblock Incorporating external data into the analysis of clinical trials via
  {Bayesian} additive regression trees.
\newblock {\em Statistics in Medicine}, 40(28):6421--6442.

\end{thebibliography}

\newpage
\begin{table}[h!]
	\centering
	\begin{tabular}{ l l }
		\hline 
		$n$
		& $\eta$ \\
		\hline
		(20, 40, 60, 80, 100, 120, 160, 200, 300, 400, 500, 600, 800, 1000)& 0\\
		(20, 40, 60, 80, 100) & -1.03\\
		(40, 80, 120, 160, 200)  & -1.36\\
		(100, 200, 300, 400, 500) &-0.56\\
		(100, 200, 300, 400, 500) &-0.83\\
		(200, 400, 600, 800, 1000)  & -0.39\\
		\hline
	\end{tabular}
	\caption{Simulation scenarios arising from sample size and effect parameter value combinations.}
	\label{table1}
\end{table}

\begin{table}[h!]
	\centering
	\begin{tabular}{ l l }
		\hline 
		$n$
		& $\eta$ \\
		\hline
		(20, 60, 100) & -1.24\\
		(40, 120, 200)  & -0.88\\
		(100, 300, 500) &-0.56\\
		(200, 600, 1000)  & -0.39\\
		\hline
	\end{tabular}
	\caption{Simulation scenarios in the training set to assess prediction performance for power.}
	\label{table2}
\end{table}

%
%

%
%
%

%
%
%
%
\newpage
\appendix
\renewcommand{\thefigure}{A\arabic{figure}}
\setcounter{figure}{0}
\section{Asymptotic validity of the models} 
\subsection{Symmetry of the sampling distribution under the null}
\label{sec:nullSymmetry}
The symmetry of the sampling distribution under the null is equivalent to the assumption that the posterior median is an unbiased estimator for $\psi$, whose true value under the null is assumed to be $\psi_0$. This assumption is reasonable in many cases including the examples of the paper where the effect parameter is obtained from a set of regression coefficients and for moderate sample sizes. In cases where this assumption is violated, it is recommended to allow the shape and scale parameters of the beta distribution to be estimated independently in (\ref{eq:betaH0}).
The equivalence of the unbiasedness of the posterior median and the symmetry of the sampling distribution under the null (i.e., ${\rm E}(\tau) = 0.5$) is shown below.

Let $\tilde{\psi}$ be the posterior median, assumed to be an unbiased estimator for $\psi$, whose true value under the null is $\psi_0$. Then,
\begin{align*}
	{\rm E}(\tau) &= {\rm E}\left(P(\psi>\psi_0\mid \mathbf{y})\right) \\
	&=  {\rm E}\left(P(\psi>\tilde{\psi}\mid \mathbf{y}) + \int_{\psi_0}^{\tilde{\psi}}\pi(\psi\mid \mathbf{y})d\psi\right)\\
	&= 0.5.
	\end{align*}

\subsection{Convergence of the moments under the null and alternative hypotheses}
\label{sec:asympmoments}
Below, we show that the proposed distributions in (\ref{eq:betaH0})-(\ref{eq:aHA}) satisfy the asymptotic behavior of the sampling distribution. To approximate the means and variances of the functions that arise in the derivations below, we use the Taylor expansion of various differentiable functions of a random variable around its mean. Specifically, let $Z$ be a random variable with ${\rm E}(Z)=\mu$ and ${\rm Var}(Z)=\sigma^2$, then informally,
$$g(Z)\approx g(\mu) + g'(\mu)(Z-\mu),$$
therefore,
$${\rm E}(g(Z))\approx g(\mu),$$
and
$${\rm Var}(g(Z))\approx [g'(\mu)]^2{\rm Var}(Z).$$

Using these approximations, we show that the specifications in (\ref{eq:betaH0}) and (\ref{eq:aH0}) satisfy the following under the null hypothesis, $ {\rm E}(\tau) = \frac{1}{2}$ and $\lim_{n\rightarrow \infty} {\rm Var}(\tau) = \frac{1}{12}$; therefore, the sampling distribution is asymptotically uniform under this specification. Similarly, we show that under the alternative hypothesis the proposed models (\ref{eq:betaHA}) and (\ref{eq:aHA}) guarantee that $\lim_{n\rightarrow \infty} {\rm E}(\tau) = 1$ and $\lim_{n\rightarrow \infty} {\rm Var}(\tau) = 0$, i.e., the sampling distribution converges to a point mass at 1.

It is straightforward to see that (\ref{eq:betaH0}) and (\ref{eq:aH0}) impose symmetry for the distribution under the null and give,
\begin{align*}
	{\rm E}(\tau) ={\rm E}\left({\rm E}(\tau \mid a_0)\right)= {\rm E}\left(\frac{a_0}{2a_0}\right) = \frac{1}{2}\\
\end{align*}
As for the variance, 
\begin{align*}
	{\rm Var}(\tau) &={\rm E}\left(\rm{V}(\tau \mid a_0)\right) + \rm{Var}\left(\rm{E}(\tau \mid a_0)\right) \\
	& =\rm{E}\left(\frac{1}{4(2a_0+1)}\right) \\
\end{align*}
Let $Z_0= \log(a_0) \sim \mathcal{N}(\frac{\alpha_1}{n} + \frac{\alpha_2}{n^2}, \sigma^2_0)$ and $g_1(Z_0)=\frac{1}{4(2\exp{(Z_0)}+1)}$, then,
\begin{align*}
	\lim_{n\rightarrow \infty} {\rm Var}(\tau) &= \lim_{n\rightarrow \infty} {\rm E}\left(g_1(Z_0)\right) \\
	&\approx \lim_{n\rightarrow \infty} g_1(\frac{\alpha_1}{n} + \frac{\alpha_2}{n^2})\\
	& = \frac{1}{12}.
	\end{align*}
	
Under the alternative,
\begin{align*}
	{\rm E}(\tau) ={\rm E}\left({\rm E}(\tau \mid a_A)\right)= {\rm E}\left(\frac{a_A^2}{a_A^2+1}\right).
\end{align*}
Let $Z_A= \log(a_A) \sim \mathcal{N}\left(\phi_1\sqrt{n}(\psi^* - \psi_0) + \phi_2 n(\psi^* - \psi_0)^2, \sigma^2_1\right)$ and $g_2(Z_A)=\frac{\exp(2Z_A)}{\exp(2Z_A)+1}$, then,
	
\begin{align*}
		\lim_{n\rightarrow \infty} \rm{E}(\tau) &=\lim_{n\rightarrow \infty}\rm{E}\left(g_2(Z_A)\right)\\ &\approx\lim_{n\rightarrow \infty}g_2(\phi_1\sqrt{n}(\psi^* - \psi_0) + \phi_2 n(\psi^* - \psi_0)^2)\\
		&\approx 1.
\end{align*}
And for the variance,
\begin{align*}
	{\rm Var}(\tau) &= {\rm Var}\left({\rm E}(\tau \mid a_A)\right)+ {\rm E}\left({\rm V}(\tau \mid a_A)\right) +\\
	& ={\rm Var}\left(\frac{a_A^2}{a_A^2+1}\right) +  {\rm E}\left(\frac{1}{(a_A + 1/a_A)^2(a_A + 1/a_A+1)}\right) 
\end{align*}
Let $g_3(Z_A) = \frac{1}{(\exp(Z_A) + 1/\exp(Z_A) )^2(\exp(Z_A)  + 1/\exp(Z_A) +1)}$, then,
given that $g'_2(Z_A) = \frac{2\exp(2Z_A)}{(\exp(2Z_A)+1)^2}$, we have,
\begin{align*}
\lim_{n\rightarrow \infty}	{\rm Var}(\tau) \approx & \lim_{n\rightarrow \infty} \left[g'_2\left(\phi_1\sqrt{n}(\psi^* - \psi_0) + \phi_2 n(\psi^* - \psi_0)^2\right)\right]^2\sigma_1^2 \\
& +  \lim_{n\rightarrow \infty} g_3\left(\phi_1\sqrt{n}(\psi^* - \psi_0) + \phi_2 n(\psi^* - \psi_0)^2\right)\\ = & 0.
\end{align*}

\subsection{Power curve estimates} 

\begin{figure}[h!]
	
	\begin{subfigure}[b]{0.47\textwidth}
		\centering
		\includegraphics[width=\textwidth]{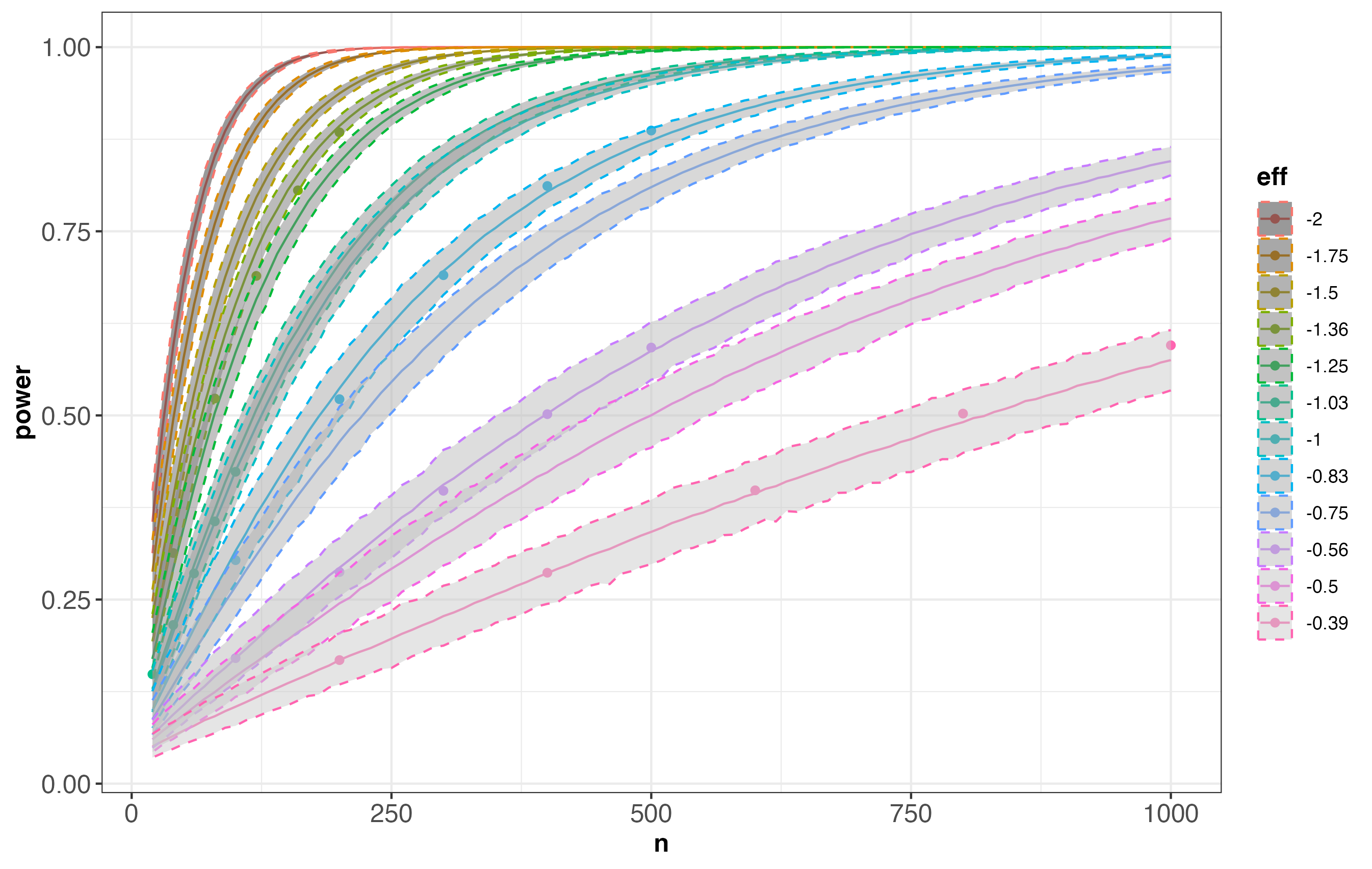}
		\caption{}
		\label{fig:power_all_unadj_col}
	\end{subfigure}
	\begin{subfigure}[b]{0.47\textwidth}
		\centering
		\includegraphics[width=\textwidth]{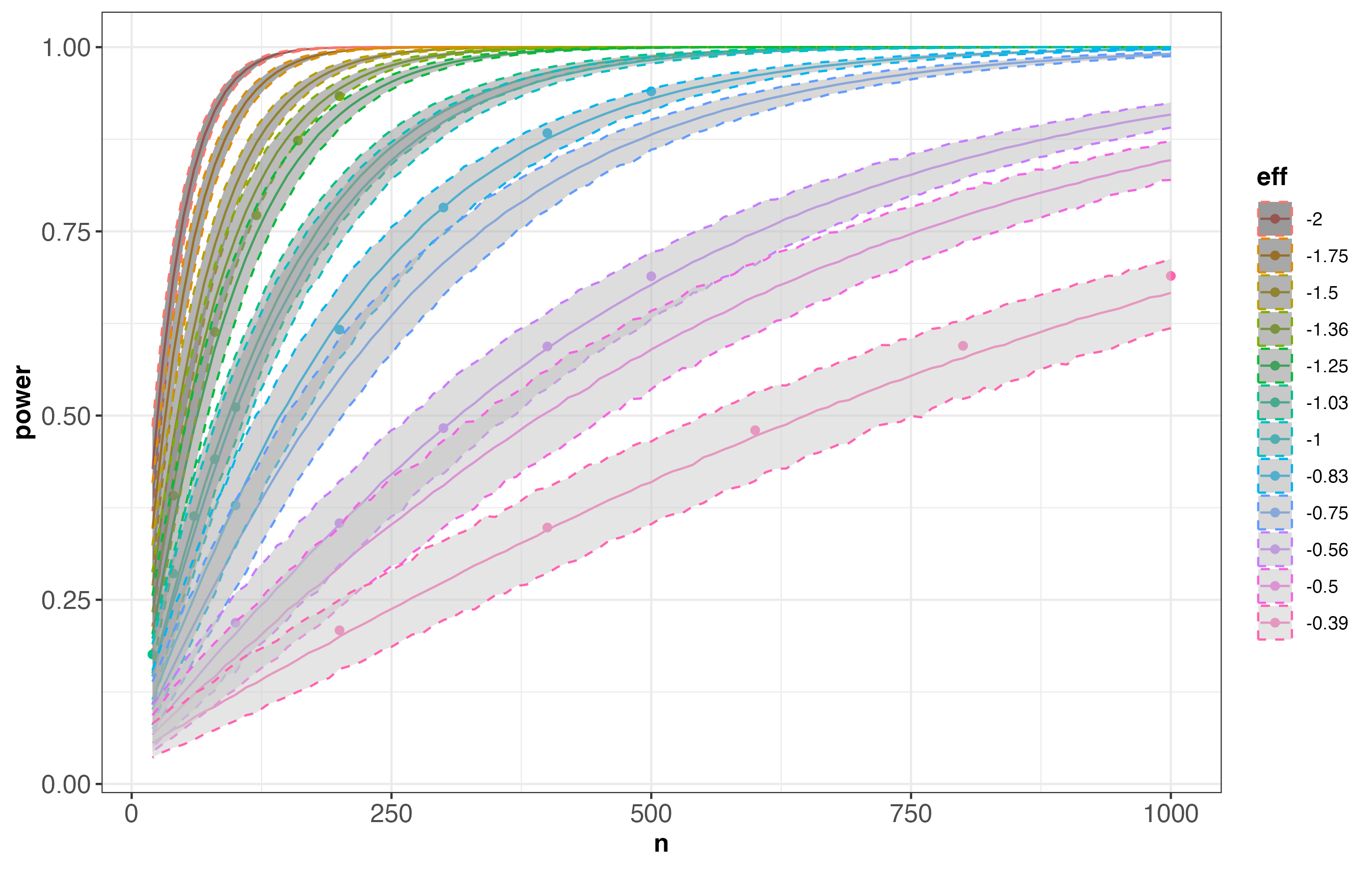}
		\caption{}
		\label{fig:power_all_adj_col}
	\end{subfigure}
	
	\caption{Estimated curves (posterior median) and 95\% credible intervals (equal-tailed posterior quantiles) for power as a function of sample size for a range of effect sizes in the (a) unadjusted and (b) adjusted model. The dots are simulation-based estimates of power at selected points.}
	\label{fig:power_col}
\end{figure}

Figure~\ref{fig:power_col} presents the results where power curves are obtained from the model for the same effect sizes as those used in the simulations, which shows that the estimates capture the simulation-generated values.

\end{document}